\documentclass[authoryear,5p]{elsarticle}
\usepackage{natbib}
\usepackage{epsfig}
\usepackage{amssymb}

\newcommand{\be}{\begin{equation}}
	
\newcommand{\ee}{\end{equation}}
\newcommand{\bea}{\begin{eqnarray}}
\newcommand{\eea}{\end{eqnarray}}
\newcommand{\bef}{\begin{figure}}
\newcommand{\eef}{\end{figure}}
\newcommand{\bce}{\begin{center}}
\newcommand{\ece}{\end{center}}
\def\lsim{\mathrel{\rlap{\lower4pt\hbox{\hskip1pt$\sim$}}
	    \raise1pt\hbox{$<$}}}         
\def\gsim{\mathrel{\rlap{\lower4pt\hbox{\hskip1pt$\sim$}}
	        \raise1pt\hbox{$>$}}}         

\begin{document}

\begin{frontmatter}

\title{Millisecond radio pulsars with known masses: parameter values and equation of state models}

\author[1]{Sudip Bhattacharyya}
\ead{sudip@tifr.res.in}

\author[2,3]{Ignazio Bombaci}
\author[4]{Debades Bandyopadhyay}
\author[5,6]{Arun V. Thampan}
\author[3]{Domenico Logoteta}

\address[1]{Department of Astronomy and Astrophysics, Tata Institute of Fundamental Research, Mumbai 400005, India}
\address[2]{Dipartimento di Fisica, Universit\`{a} di Pisa, Largo B. Pontecorvo, 3 I-56127 Pisa, Italy}
\address[3]{INFN, Sezione di Pisa, Largo B. Pontecorvo, 3 I-56127 Pisa, Italy}
\address[4]{Astroparticle Physics and Cosmology Division, Saha Institute of Nuclear Physics, HBNI, 1 / AF Bidhannagar, Kolkata-700064, India}
\address[5]{Department of Physics, St. Joseph's College, 36 Lalbagh Road, Bangalore 560027, India}
\address[6]{Inter-University Centre for Astronomy and Astrophysics (IUCAA), Post Bag 4, Ganeshkhind, Savitribai Phule Pune University Campus, Pune 411007, India}





\begin{abstract}
The recent fast growth of a population of millisecond pulsars with precisely measured mass
provides an excellent opportunity to characterize these compact stars at an unprecedented level.
This is because the stellar parameter values can be accurately computed
for known mass and spin rate and an assumed equation of state (EoS) model.
For each of the 16 such pulsars and for a set of EoS models from
nucleonic, hyperonic, strange quark matter and hybrid classes, we numerically compute fast spinning
stable stellar parameter values considering the full effect of general relativity.
This first detailed catalogue of the computed parameter values of observed millisecond pulsars provides a testbed
to probe the physics of compact stars, including their formation, evolution and EoS.
We estimate uncertainties on these computed values from the uncertainty of the measured mass,
which could be useful to quantitatively constrain EoS models. We note that
the largest value of the central density $\rho_{\rm c}$ in our catalogue is $\sim 5.8$ 
times the nuclear saturation density $\rho_{\rm sat}$, which is much less than the expected maximum 
value $13 \rho_{\rm sat}$. We argue that the $\rho_{\rm c}$-values of at most a small fraction of 
compact stars could be much larger than $5.8 \rho_{\rm sat}$.
Besides, we find that the constraints on EoS models from accurate radius measurements could be significantly 
biased for some of our pulsars, if stellar {\it spinning} configurations are not used to compute the 
theoretical radius values.
\end{abstract}

\begin{keyword}
equation of state \sep  methods: numerical \sep pulsars \sep stars: neutron \sep stars: rotation
\end{keyword}

\end{frontmatter}



\section{Introduction}\label{Introduction}

Compact stars, commonly known as ``neutron stars", are possibly the densest objects in the universe apart
from black holes. A class of such compact stars, pulsars, show periodic variation of intensity 
in their electromagnetic emission.  
In fact, compact stars were discovered from such periodic radio pulses \citep{Hewishetal1968}.
The first reported fast spinning pulsar, i.e., the millisecond (ms) pulsar, was PSR B1937+21
\citep{Backeretal1982}. It was immediately proposed that such pulsars could be spun up
by accretion-induced angular momentum transfer in low-mass X-ray
binaries \citep[LMXBs;][]{RadhakrishnanSrinivasan1982, Alparetal1982}.
This model was strengthened by the discovery of an accretion-powered X-ray ms pulsar 
SAX J1808.4-3658 \citep{WijnandsKlis1998, ChakrabartyMorgan1998}. This is because
this X-ray pulsar showed that compact stars could be spun up in LMXBs. However,
a clear evolutionary connection between radio ms pulsars and LMXBs would be established
if a compact star had shown both LMXB phase and radio pulsations
possibly in a non-accreting phase. Recently, three such sources, called 
transitional pulsars, have been discovered \citep{Archibaldetal2009, Papittoetal2013, deMartinoetal2013}.
These findings strongly show that ms pulsars are spun up in LMXBs.
However, the detailed mechanism of this spin evolution, which depends on
accretion processes and the interaction between the accretion disc and the stellar magnetosphere,
is somewhat poorly understood. Testing the models of these physical processes against the
precisely measured parameter values of observed ms pulsars will be very useful to
understand the physics of compact star evolution.

Another poorly understood aspect of compact stars is their internal composition, especially 
the physics of their cores. The densities of these degenerate cores are well above the nuclear 
saturation density $\rho_{\rm sat} \approx 2.6\times10^{14}$~g cm$^{-3}$. 
Consequently \citep[see, e.g.,][]{bombaci2007} various particle species 
(apart from neutrons, protons, electrons and muons) 
and phases of dense matter are expected in the stellar interior.  
Thus different types of compact stars (nucleonic, hyperonic, strange matter, hybrid) are hypothesized 
to exist. Therefore compact stars can be considered as natural laboratories that allow us to investigate the  constituents of matter and their interactions under extreme conditions that cannot be reproduced 
in any terrestrial laboratory.
Understanding the nature of supra-nuclear core matter remains a fundamental problem of physics, even after almost 
50 years since the discovery of the first pulsar. 

The standard way attempted to solve this problem is the following. 
Assuming the constituents of the stellar matter and the interactions among them, an equation of state (EoS) 
is computed using different many-body approaches.
The EoS is given by the thermodynamical relation between the matter pressure $P$, the mass density $\rho$  
and the temperature $T$.  The temperature could be considered equal to zero a few minutes after the compact star  
birth \citep{BurLat86,bomb+95,prak97}.
Many such EoS models exist in the literature.


In order to understand the superdense matter of compact star cores, it is required to 
identify the ``correct" EoS model. How can one do that? For a proposed EoS model, one
can compute the stable stellar structure. For doing this, one needs to solve the
Tolman-Oppenheimer-Volkoff (TOV) equations \citep{OppenheimerVolkoff1939, Tolman1939}
for nonspinning compact stars. For fast spinning stars, however, one needs to follow 
a numerical formalism described and used in this paper. Once the stable stellar structure
is computed, it is possible to calculate the values of various stellar parameters, such
as mass, radius, spin rate, etc. One needs to compare these computed values with the 
measured values to reject some proposed EoS models. By rejection of many EoS models,
one can attempt to identify the ``correct" EoS model as accurately as possible.
In order to achieve this goal, one needs to measure three independent parameters of 
the same ms pulsar.

So far, for no compact star three parameters have been precisely measured.
However, precise measurements of two parameters, mass and spin rate, have been
done for a fast growing population of ms pulsars in recent years. This, for the
first time, provides a unique opportunity to characterize a number of observed ms pulsars 
with an unprecedented accuracy. Note that
previous authors explored the stable structure of compact stars, which provided general
information about the fast spinning compact star parameters \citep[e.g.,][]{Cooketal1994}.
Some authors went a step further, considered the measured spin rate of an ms pulsar, and computed a constant
spin sequence for that pulsar \citep[e.g.,][]{Dattaetal1998,Bhattacharyyaetal2016}. But
such a sequence gives large ranges of other parameter values for a given EoS model.
Therefore, while this sequence gives an insight about the general properties of
ms pulsars, it does not give very useful additional information about the parameters of the considered
ms pulsar. Furthermore, since these large ranges of parameter values overlap for various EoS
models, such spin sequences are not very useful to constrain EoS models.
With two parameters known, we can now accurately estimate the other parameter values of 
observed ms pulsars for a given EoS model. In this paper, we do this estimation for
16 ms pulsars and eight diverse EoS models from four different classes, and make a catalogue. This catalogue
not only will be useful to constrain EoS models, but also will provide a testbed to probe
the physical processes of compact star evolution. We also discuss the ways to constrain 
EoS models using measured radius values.

In \S~\ref{EoS}, we mention and discuss the ms pulsars and EoS models we consider.
In \S~\ref{Computation}, we describe the procedure to compute stable fast spinning compact
star structure. \S~\ref{Results} includes our catalogue of computed ms pulsar parameters
and a detailed discussion on their implications. 
In \S~\ref{Summary}, we summarize our results and conclusions.

\section{Pulsars and equations of state}\label{EoS}

As mentioned in \S~\ref{Introduction}, in this work we use a special sample of ms pulsars, 
that is those with precisely measured mass values. 
Here we define pulsars with spin periods less than 10 ms as ms pulsars.
Masses of these pulsars in binary stellar systems were measured 
from the estimation of post-Keplerian parameters or the spectroscopic observations 
of the companion white dwarf stars (see \citet{OzelFreire2016} and references therein).
We choose ms pulsars with quoted mass error less than a quarter of a
solar mass. 
We list the spin rates, measured 
masses with errors and the references on spin and mass measurements for these pulsars
in Table~\ref{table_pulsar}. It is interesting to see that masses of all these pulsars
are distributed within the $1-2 M_\odot$ range. 

\begin{figure*}
\centering
\hspace{-0.8cm}
\includegraphics*[width=15cm]{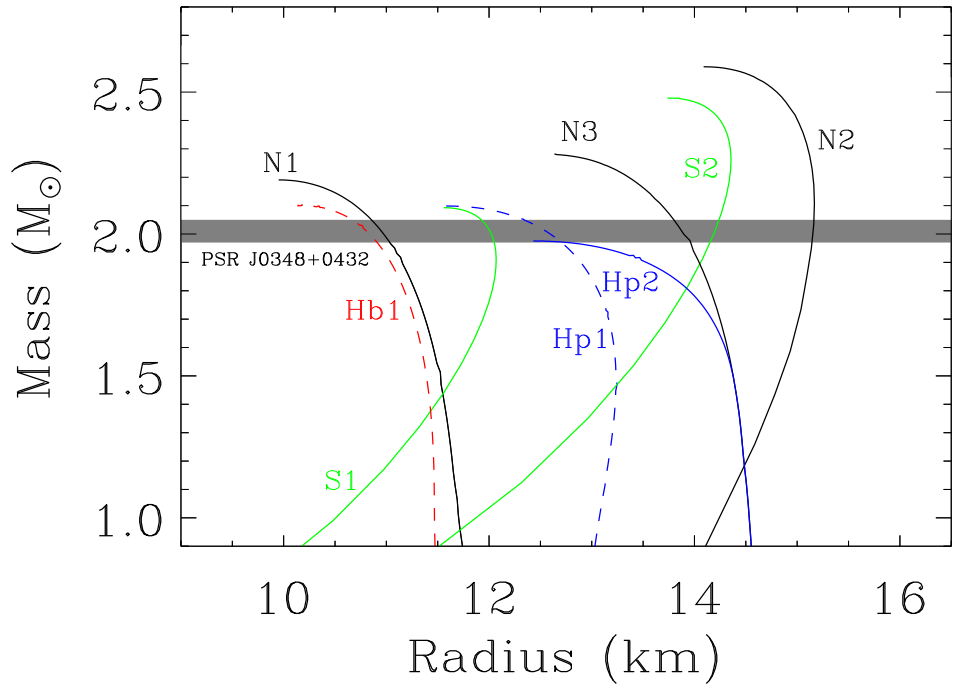}
\caption{Compact star gravitational mass versus radius curves. Eight curves
	for non-spinning stars (\S~\ref{Computation})
	are for eight EoS models (marked by names given in Table~\ref{table_EoS}).
	The horizontal band shows the mass ($2.01\pm0.04 M_\odot$) 
	of the relatively slowly spinning (spin period $= 39$ ms) pulsar PSR J0348+0432.
	This figure shows that all our EoS models can support the mass of this massive pulsar (see \S~\ref{EoS}).
\label{fig1}}
\end{figure*}

In this paper, we consider eight EoS models from four different classes 
(Table~\ref{table_EoS}). We carefully chose these EoS models keeping various points in
mind. For example, three of our EoS models are nucleonic, two are hyperonic,
two are strange quark matter and one is hybrid, and hence the set is truly diverse.
Moreover, the discovery of the massive pulsar PSR J0348+0432 with a precisely
measured mass ($2.01\pm0.04 M_\odot$; \citet{Antoniadisetal2013}) demands that
the ``correct" EoS must be able to support this high mass. All our EoS models
pass this test (Figs.~\ref{fig1}, \ref{fig2} and Table~\ref{table_EoS}). We also note that the
discoveries of this pulsar and another massive pulsar (PSR J1614-2230; 
\citet{Demorestetal2010}; see also Table~\ref{table_pulsar})
essentially constrained the radius space from the lower side. This is because, 
within a certain class of EoS models (e.g., nucleonic or strange matter),
a harder EoS  
which can support a higher maximum mass gives a higher radius value for a given mass (see Fig.~\ref{fig1}). 
Therefore, as indicated by this figure, EoS models with radius values less than a certain limit in 
the observed $1-2 M_\odot$ mass range (Table~\ref{table_pulsar}) may not be able to support the high mass 
values of PSR J0348+0432 and PSR J1614-2230, thus effectively shrinking the radius space. 
Our EoS models nicely fill this shrunken radius space (see Fig.~\ref{fig1}). 

Here we give a brief description for each EoS model.\\ 
(1) Nucleonic matter: The first nucleonic EoS model (N1), denoted by A18+$\delta v$+UIX (\citet{Akmaletal1998}; 
Table~\ref{table_EoS}), is based on the Argonne $v_{18}$ model A18 \citep{Wiringaetal1995} 
of two-nucleon interaction. 
The A18 model fits very well the phase shifts for nucleon-nucleon scattering of the Nijmegen database 
\citep{Stoksetal1993}. The A18+$\delta v$+UIX model additionally includes three-nucleon interactions described 
by the Urbana IX [UIX] model \citep{Pudlineretal1995} and the effect of relativistic boost corrections. 
The second nucleonic EoS model (N2; \citet{Sahuetal1993}; Table~\ref{table_EoS}), 
which is harder (\S~\ref{Introduction}) than N1, is a field theoretical EoS for nucleonic matter  
in $\beta$-equilibrium based on the chiral sigma model. The model includes an isoscalar vector field 
generated dynamically. 
The third nucleonic EoS model (N3; \citet{SugaharaToki1994, providencia2013}; Table~\ref{table_EoS}) 
is based on a relativistic mean field (RMF) approach in which nucleons ($n$, $p$) interact via 
the exchange of $\sigma$, $\omega$ and $\rho$ mesons. In particular, in the present work we use 
the parameters set denoted as TM1-2 in Table I of \citet{providencia2013}.  
All the nucleonic EoS models used in our calculations reproduce the empirical saturation point 
of nuclear matter $n_{sat} = 0.16 \pm 0.01~{\rm fm}^{-3}$,  $E/A|_{n_{sat}} = -16.0 \pm 1.0~{\rm MeV}$ 
\citep[e.g.,][]{Logoteta2015} and the empirical value of the nuclear symmetry energy 
$E_{sym}(n_{sat}) = 28$ -- $33$~MeV, at saturation density.
\\ 
(2) Hyperonic matter: The first hyperonic EoS model \citep[Hp1;][]{Baniketal2014} 
is based on a RMF approach in which nucleons and $\Lambda$ hyperons interact via the exchange of $\sigma$, 
$\omega$ and $\rho$ mesons with the additional contribution of the hidden-strangeness meson $\phi(1020)$ 
and using density dependent coupling constant.
The second hyperonic EoS model \citep[Hp2;][]{providencia2013} is also based on a RMF approach, 
which includes all the members of the $J^\pi = (1/2)^+$ baryon octet  
(i.e. $n$, $p$, $\Lambda$, $\Sigma^-$, $\Sigma^0$, $\Sigma^+$, $\Xi^-$ and $\Xi^0$)  
interacting via $\sigma$, $\omega$, $\rho$ and hidden-strangeness $\sigma^*$ and $\phi(1020)$ meson 
exchange. Among the different hyperonic EoS parametrizations reported in \citet{providencia2013},
in the present work we use the one corresponding to the TM1-2 parameters set for the nucleonic sector,
without $\sigma^*$ mesons, with $\Lambda_{\omega} = 0$ and taking for the potential energy depths  
for the $\Lambda$, $\Sigma$, and $\Xi$ hyperons in symmetric nuclear matter at saturation density 
$n_{\rm sat}$ with the values $U_{\Lambda} = -28~{\rm MeV}$, $U_{\Sigma} = 30~{\rm MeV}$, 
$U_{\Xi} = 18~{\rm MeV}$ respectively \citep[see Table II in][]{providencia2013}. 
\\
(3) Strange quark matter: These EoS models are the simple version of the MIT
bag model, which was extended to include perturbative corrections due to quark interactions, 
up to the second order in the strong structure constant \citep{Fra01, Alf05, weis11}. These
EoS models are characterized with two parameters: effective bag constant ($B_{\rm eff}$) and
perturbative QCD corrections term parameter ($a_4$). The value $a_4 = 1$ corresponds to the ideal
relativistic Fermi gas EoS. For the first model (S1),
$B_{\rm eff}^{1/4} = 138$ MeV, $a_4 = 0.8$, while for the second model (S2),
$B_{\rm eff}^{1/4} = 125$ MeV, $a_4 = 0.5$ \citep{Bhattacharyyaetal2016}.
Note that S2 is harder (\S~\ref{Introduction}) than S1.
\\
(4) Hybrid (nuclear+quark) matter: In this class of models, one assumes the occurence of 
the quark deconfinement phase transition in the neutron star core.   
Following \citet{glendenning1992,glendenning1996}, we model the nuclear to quark matter 
transition as a first order phase transition occuring in a multicomponent system with two conserved 
``charges" (the electric charge and the baryon number). The specific hybrid star matter EoS model (Hb1)   
considered in the present paper, has been obtained using the A18+$\delta v$+UIX EoS \citep{Akmaletal1998} 
for the nuclear matter phase and the extended MIT bag model EoS \citep{Fra01, Alf05, weis11} 
for the quark phase with $B_{\rm eff}^{1/4} = 138$ MeV, $a_4 = 0.4$. 

\section{Fast spinning stellar structure computation}\label{Computation}

Here we briefly mention the method to compute fast spinning stable compact star 
structures, the corresponding stellar parameters and the equilibrium sequences. 
A detailed description of the method can be found in \citet{Cooketal1994}.
Such computation requires a general relativistic treatment.
The general spacetime of such a star is (using $c=G=1$; \citet{Bardeen1970, Cooketal1994}):
\begin{eqnarray}
{\rm d}s^2 = -{\rm e}^{\gamma+\rho}{\rm d}t^2 + {\rm e}^{2\alpha}({\rm d}r^2 + 
r^2{\rm d}\theta^2) + {\rm e}^{\gamma-\rho}r^2\sin^2\theta \nonumber \\
({\rm d}\phi - \omega{\rm d}t)^2,
\end{eqnarray}
where $t$, $r$ and $\theta$ are temporal, quasi-isotropic radial and polar angular
coordinates respectively,
$\gamma$, $\rho$, $\alpha$ are metric potentials, and $\omega$
is the angular speed of the stellar fluid relative to the local inertial frame.
Einstein's field equations are solved to compute the $r$ and $\theta$
dependent $\gamma$, $\rho$, $\alpha$ and $\omega$, as well as the stable stellar structure,
for a given EoS model, and assumed values of two parameters, such as stellar
central density ($\rho_{\rm c}$) and polar radius to equatorial radius ratio
\citep{Cooketal1994, Dattaetal1998, Bombacietal2000, 
Bhattacharyyaetal2000, BhattacharyyaBhattacharyaThampan2001, 
BhattacharyyaMisraThampan2001, BhattacharyyaThampanBombaci2001, Bhattacharyya2002,
Bhattacharyya2011}.
This equilibrium solution is then used to compute
compact star parameters, such as gravitational mass ($M_{\rm G}$), rest
mass ($M_{\rm 0}$), equatorial circumferential radius ($R_{\rm e}$),
spin frequency ($\nu$), total angular momentum ($J$), moment of inertia ($I$),
total spinning kinetic energy ($T$) and total gravitational energy ($W$)
\citep{Cooketal1994, Dattaetal1998}.

The radius $r_{\rm ISCO}$ of the innermost stable circular orbit (ISCO)
is calculated in the following way. The radial equation of motion
around such a compact star is
\.{r}$^2 \equiv {\rm e}^{2\alpha + \gamma + \rho}({\rm d}r/{\rm d}\tau)^2 = $\~{E}$^2 - 
$\~{V}$^2$, where, ${\rm d}\tau$ is the proper time, \~{E} is the specific
energy, which is a constant of motion, and \~{V} is the effective potential.
The effective potential is given by \~{V}$^2 = {\rm e}^{\gamma+\rho}[1 + 
\frac{l^2/r^2}{{\rm e}^{\gamma-\rho}}] + 2\omega$\~{E}$l - \omega^2 l^2$.
Here $l$ is the specific angular momentum and a constant of motion.
We determine $r_{\rm ISCO}$ using the condition \~{V}$_{,rr}$ = 0,
where a comma followed by one $r$ represents a first-order
partial derivative with respect to $r$ and so on \citep{ThampanDatta1998}.

We compute the static or nonspinning limit, where $\nu \rightarrow 0$
and $J \rightarrow 0$, for all EoS models (see Figs.~\ref{fig1} and \ref{fig2}). 
The main aim of this paper is to compute various parameter
values of all the 16 pulsars mentioned in Table~\ref{table_pulsar}. 
We obtain the stable configuration for known mass and spin rate of each ms pulsar and for each 
EoS model using multi-iteration runs of our numerical code.
Such multi-iteration runs include computations of the constant $\nu$ equilibrium sequence.
Then various parameter values describing this equilibrium configuration are obtained.
It is also useful to obtain the uncertainties in these parameters. Note that
measurement errors of $\nu$ values are sufficiently small (e.g., spin period
$\sim 5.7574518191(9)$~ms for PSR J0437-4715; \citet{Johnstonetal1993}).
So the uncertainties of the computed parameters essentially come from the 
measurement errors of stellar mass values. These errors of $M_{\rm G}$ for each ms pulsar,
as quoted in Table~\ref{table_pulsar}, give  lower ($M_{\rm G}^{\rm l}$) and
upper ($M_{\rm G}^{\rm u}$) mass values. We compute the two limits
of each computed parameter for each pulsar and EoS model using 
$M_{\rm G} = M_{\rm G}^{\rm l}$ and $M_{\rm G} = M_{\rm G}^{\rm u}$ and the
measured $\nu$. These two limits for a parameter give the uncertainties 
quoted in Table~\ref{table_result}.

\section{Results and discussion}\label{Results}

\subsection{Properties of millisecond pulsars}\label{Properties}

Here we present our results in Table~\ref{table_result}, and discuss their implications.
This table displays a catalogue of numerically computed parameter values of 16 ms pulsars (from Table~\ref{table_pulsar}) with 
precisely measured mass. 
For each pulsar, we calculate parameter values with error bars for each of eight EoS models 
(Table~\ref{table_EoS}), using the procedure mentioned in \S~\ref{Computation}.

Since we compute stellar parameter values for diverse 
EoS models, the numbers given in Table~\ref{table_result} characterize observed ms pulsars at an unprecedented level. 
Moreover, since our sample pulsars have diverse mass and spin values, they may be representative enough for using this
knowledge to understand other compact stars, including some of the fast spinning accreting stars in LMXBs.
Therefore, the catalogue will provide a unique testbed to probe the physics of compact stars, including
their formation, evolution and EoS, specifically for the 16 ms pulsars (Table~\ref{table_result}). 
Below we discuss what we learn about a number of properties of these 
16 ms pulsars, and their implications, based on detailed 
general relativistic computation of their structures using their measured masses and spin rates, and
realistic EoS models.

\noindent
{\bf (1) Central density ($\rho_{\rm c}$):}

The central part of a compact star harbours possibly the densest matter in the universe, which
is not hiding behind an event horizon. So it is tantalizing to know how dense this matter can be 
for observed compact stars, especially those with measured mass values. 
The knowledge of the range of this maximum density for compact stars with known masses should have
impact on our understanding of constituents and physics of the dense matter. Such knowledge will
also characterize the compact star populations, and will be important to understand their formation
and evolution. This is because the $\rho_{\rm c}$-value at the birth is expected to depend on the
astrophysical process related to the compact star formation. Moreover, this initial $\rho_{\rm c}$ evolves
into our inferred $\rho_{\rm c}$-value, because ms pulsars acquire mass and angular 
momentum during their LMXB phases \citep[e.g.,][]{Bejgeretal2011}.

The issue of the largest possible density $\rho_{\rm max}$ in compact stars has been previously investigated 
in \citet{LattimerPrakash2005}, where the authors report the calculated maximum mass ($M_{\rm G}^{\rm max}$) 
for non-spinning compact stars versus the corresponding  central density for various EoS models. 
In addition, they make the conjecture that the analytic Tolman VII solution \citep{Tolman1939} of the 
TOV equations marks the upper limit for the density reachable inside a compact star.   
The present accurate mass measurements for PSR J0348+0432 with $M = 2.01\pm0.04 M_\odot$ \citep{Antoniadisetal2013}, 
using the argument of \citet{LattimerPrakash2005} (see their Fig. 1), 
implies $\rho_{\rm max} \sim  13 \rho_{\rm sat}$. Here, $\rho_{\rm sat}$ is the nuclear saturation density
($\approx 2.6\times10^{14}$~g cm$^{-3}$).
More recently, this argument has been also discussed in \citet{LattimerPrakash2010}, where the authors discuss 
also the limiting  case of the so-called ``maximally compact EoS" \citep{HaenselZdunik1989}; 
but again they get $\rho_{\rm max} \sim  13 \rho_{\rm sat}$.

Table~\ref{table_result} displays the $\rho_{\rm c}$-ranges of 16 ms pulsars, each for eight EoS models. 
This table shows that a harder EoS model within an EoS class has a lower $\rho_{\rm c}$
for given gravitational mass ($M_{\rm G}$) and spin frequency ($\nu$) values.
This is because the interaction between the stellar constituents is more repulsive for harder 
EoS models and hence it is more difficult to compress the matter.  
Here we note that, within the hyperonic class, Hp2 is harder below a certain density, and Hp1 is
harder above this density. This could be possible because, while Hp1 includes only Lambda hyperons,
Hp2 includes both Lambda and Sigma hyperons. This is why the $\rho_{\rm c}$ value is higher for Hp1 
(compared to that for Hp2) except for the highest mass pulsar PSR J1614-2230 in Table~\ref{table_result}.
This table also shows that a more massive ms pulsar has higher $\rho_{\rm c}$ due
to larger gravitational compression (compare, for example, the numbers for PSR J1946+3417 and PSR J1911-5958A,
which have similar $\nu$ values).


Table~\ref{table_result} shows that the maximum $\rho_{\rm c}$-value for our sample of pulsars and our
sample of diverse EoS models is $15.13_{-0.32}^{+0.34} \times 10^{14}$~g cm$^{-3}$ or $\approx 5.8_{-0.1}^{+0.1} \rho_{\rm sat}$.
This value, which corresponds to the most massive pulsar (PSR~J1614-2230) and a soft EoS model Hb1 of Table~\ref{table_result},
is significantly lower than the currently believed highest 
possible $\rho_{\rm c}$-value ($\sim 13\rho_{\rm sat}$; \citet{LattimerPrakash2005,
LattimerPrakash2010}). Can the $\rho_{\rm c}$-value of a compact star be much larger 
than the maximum value we find here? We discuss this point below.

\begin{figure*}
\centering
\hspace{-0.8cm}
\includegraphics*[width=15cm]{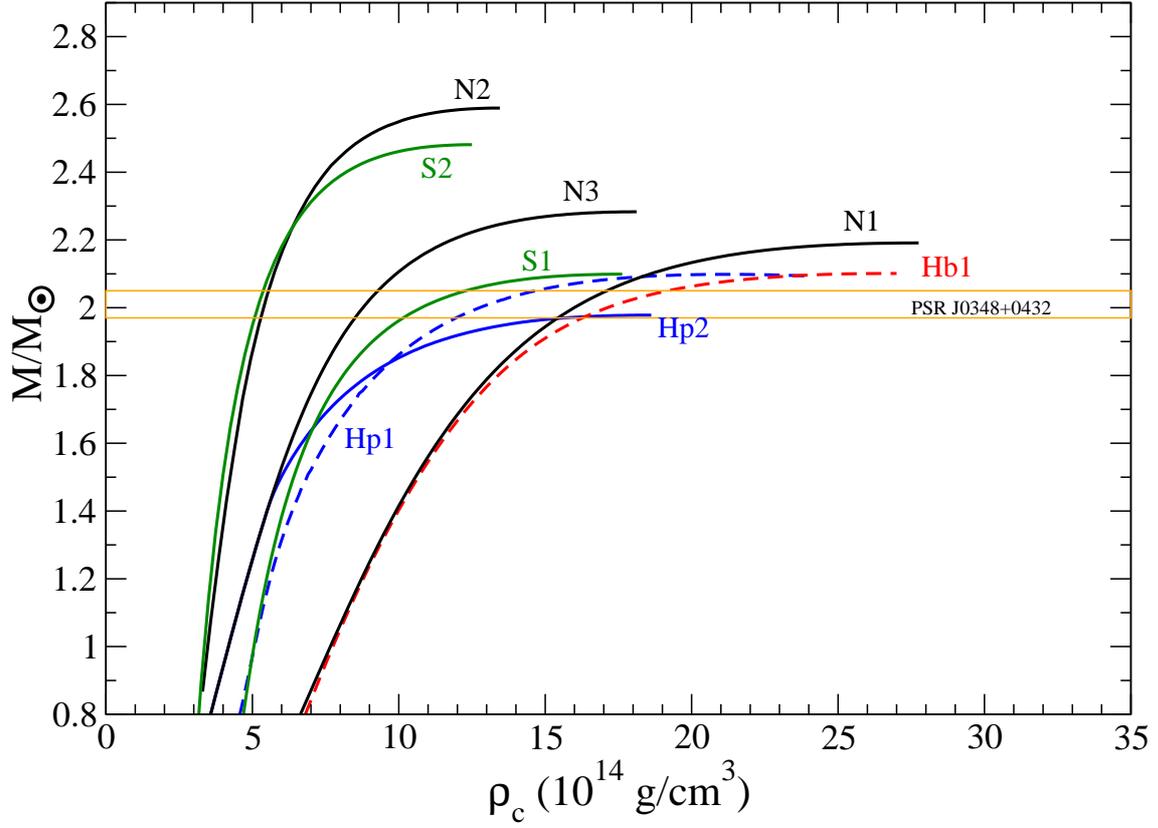}
\caption{Compact star gravitational mass versus central density curves. Eight curves
	 for non-spinning stars (\S~\ref{Computation})
	 are for eight EoS models (marked by names given in Table~\ref{table_EoS}).
	 The horizontal lines show the mass ($2.01\pm0.04 M_\odot$)
         of the relatively slowly spinning (spin period $= 39$ ms) pulsar PSR J0348+0432.
\label{fig2}}
\end{figure*}

Fig.~\ref{fig2} shows that the mass ($M_{\rm G}$) versus $\rho_{\rm c}$ curve for each EoS has a positive slope with 
two distinct parts: (1) one with high slope (almost vertical for harder EoS models),
which occupies almost the entire mass range; and (2) one with low slope (almost horizontal), in
which the $\rho_{\rm c}$-value significantly increases in a small mass range near the maximum mass
that can be supported by a given EoS model.
Now let us imagine an EoS model, which is the ``correct" EoS model. This model cannot be much softer
than the soft EoS models of our sample, because then it would not be able support the mass of the observed
most massive pulsar (PSR J0348+0432; see Fig.~\ref{fig2}). Therefore, Fig.~\ref{fig2} strongly 
suggests that the high-slope part of the $M_{\rm G}-\rho_{\rm c}$ curve
of the ``correct" EoS model cannot give a $\rho_{\rm c}$-value much larger than the maximum value
($\approx 5.8\rho_{\rm sat}$) we find here. The low-slope part for the ``correct" EoS model could, however, provide 
a much larger $\rho_{\rm c}$-value. But since the low-slope part occupies a small mass range,
only a small fraction of all compact stars, which have masses close to the maximum allowed mass,
could have $\rho_{\rm c}$-values much larger than $5.8\rho_{\rm sat}$. However, if the ``correct" EoS model
is as hard as one of our harder EoS models (e.g., N2, S2), which can be confirmed if the mass of a more
massive compact star is precisely measured in the future, then the $\rho_{\rm c}$-value of no
compact star can be as high as $5.8\rho_{\rm sat}$. Therefore, we conclude that 
the $\rho_{\rm c}$-values of at most a small fraction of compact stars could be
much larger than the maximum value we find here.

\noindent
{\bf (2) Rest mass ($M_0$):}

The rest mass (also referred to as baryonic mass) of a compact star is an important parameter.  
It can be written as $M_0 = m_u N_B$, where $m_u = 931.49$~MeV is the atomic mass unit 
and $N_B$ ($\sim$~a~few~$10^{57}$) is the total number of baryons in the star.   
For an isolated compact star, as a consequence of the baryon number conservation, $M_0$ is constant, 
whereas the corresponding values of the stellar gravitational mass $M_G$ and of the total stellar binding 
energy $B = M_0 - M_G$ depend on the EoS \citep{bombacidatta2000} and on the stellar spin frequency.  
Thus, a non-accreting compact star evolves conserving its $M_0$ value \citep{Cooketal1994}. 

The evolution of an accreting compact star depends on the binary properties and on the accretion processes 
which determine the accreted mass $\Delta M_0$ in a certain time span $\Delta t$.  
However, for a given accreted rest mass $\Delta M_0$, the increase $\Delta M_{\rm G}$ of the stellar gravitational 
mass will always be smaller than $\Delta M_0$ and will depend on the EoS. In fact, one has
\be
   \Delta M_{\rm G} =  \Delta M_0 -  \Delta B < \Delta M_0 \, , 
\ee
where $\Delta B$ is the increase of the stellar binding energy due to accretion \citep{Bagchi2011}.  
This energy can be radiated by the system during the accretion stage not only as 
electromagnetic radiation (mostly X-rays), but also via neutrino emission, since the change in the 
total stellar rest mass due to accretion alters the $\beta$-equilibrium conditions in the stellar core.


One finds from Table~\ref{table_result} that strange matter EoS models have much higher 
$B$ values than nucleonic, hyperonic and hybrid EoS models. Besides, a softer EoS model has higher 
$B$ value than a harder EoS model within a class. 
We find $B$ in the range $\sim 0.1-0.5 M_\odot$ for our sample of pulsars and EoS models.
This is $\sim 7-30\%$ of the $M_{\rm G}$ values ($\sim 6-23\%$ of the $M_0$ values) 
of our sample pulsars, which is consistent with an upper limit of 25\%
of $M_0$ reported by \citet{LattimerPrakash2010}. 
This shows that a significant fraction of the accreted matter (rest-mass energy) was lost from 
the binary system via radiation, and via neutrino emission, for our 16 ms pulsars during their LMXB phases. This shows the importance of considering such loss of accreted matter in the modeling of binary evolution 
of these ms pulsars.  

Since the effect of the total binding energy $B$ on the evolution of compact star spin rate 
and other properties is important, it is useful to have a simple relation between 
$M_{\rm G}$,  $M_0$ and $J$ to compute the stellar spin evolution. 
Such a relation was proposed by \citet{Cipollettaetal2015} in their Eq. (20).
A similar relation was earlier given by Eq. (93) of \citet{prak97} for the total binding energy
of non-spinning stars.
We check that for our nucleonic, hyperonic and hybrid EoS models, the relation by \citet{Cipollettaetal2015} gives 
values consistent with 2\% accuracy (as claimed by those authors), but the error is $\sim 6-12\%$ for our 
strange matter EoS models. This relation also usually works better for harder EoS models within a class.

\noindent
{\bf (3) Radius and oblateness:}

Attempts are being made for decades to measure the radii of compact stars using various spectroscopic
and timing methods \citep[e.g.,][]{Paradijs1978, Gendreetal2003, Bhattacharyyaetal2005, Bogdanovetal2007}. 
Such a measurement is expected to be very useful to constrain EoS models. In Table~\ref{table_result}, 
we list computed equatorial radius ($R_{\rm e}$) and polar radius ($R_{\rm p}$) values with errors of 
16 ms pulsars for each of eight EoS models.
These values can be used to constrain these models, if the radius of any of these pulsars is observationally estimated.
Table~\ref{table_result} gives a radius range of $\approx 11-16$~km for all pulsars and EoS models, 
which implies that this is roughly the radius range which one needs to constrain.
This range is consistent with previous findings \citep[e.g.,][]{Hebeleretal2013}.
Besides, note that the knowledge of oblateness of a spinning compact 
star may be important to understand its physics, and to constrain its EoS models using various
techniques \citep[e.g.,][]{MillerLamb2015, Baubocketal2013}. The $R_{\rm p}/R_{\rm e}$ value, which
determines the oblateness, decreases with increasing spin rate. For example, our fastest spinning 
pulsar has $R_{\rm p}/R_{\rm e} \approx 0.89$ for our hardest nucleonic EoS model, and our slowest
spinning pulsar has $R_{\rm p}/R_{\rm e}$ value consistent with $1.00$ for the two strange 
matter EoS models (Table~\ref{table_result}). 
This table gives an idea about typical $R_{\rm p}/R_{\rm e}$ values of our 16 ms pulsars, which can be
incorporated in techniques to constrain EoS models.

\noindent
{\bf (4) Radius-to-mass ratio ($R_{\rm e}/r_{\rm g}$):}

Measurement of stellar radius-to-mass ratio, which is the inverse of stellar compactness, 
is an alternative to radius measurement for constraining EoS models.
Here, $r_{\rm g}$ ($ = G M_{\rm G}/c^2$) is the Schwarzschild radius.
Plausible detection and identification of an atomic spectral line from the stellar surface
can provide the cleanest way to measure this parameter, even when the line is broad and skewed due to
spin-induced Doppler effect \citep{Bhattacharyyaetal2006}. This is
because the surface gravitational redshift depends on this parameter.
The intensity variation due to one or more hot spots on the spinning stellar surface
can also constrain $R_{\rm e}/r_{\rm g}$ \citep{Bhattacharyyaetal2005}. Table~\ref{table_result}
shows that this parameter has a larger value for harder EoS models within a class.
This value of the dimensionless $R_{\rm e}/r_{\rm g}$ is in the range $\approx 3.9-8.4$
for all 16 ms pulsars and eight EoS models of our sample (Table~\ref{table_result}). 
This suggests that these compact stars are not compact enough, i.e., $R_{\rm e}/r_{\rm g} > 3.5$.
A photon emitted from the stellar surface is deflected by more than $180^{\rm o}$ in the Schwarzschild spacetime
\citep{Pechenicketal1983} for $R_{\rm e}/r_{\rm g} < 3.5$, which implies multiple paths of photons from the 
surface to the observer. Such multiple paths can make the numerical computation of ray tracing,
which is required to model the spectral and timing features of the stellar surface and hence to constrain 
EoS models \citep{Bhattacharyyaetal2005}, substantially more complex. Our finding suggests that such complex 
modeling may not be required for our sample of pulsars, and in view of the diversity of our mass and
EoS samples, possibly for most compact stars.

\noindent
{\bf (5) Innermost stable circular orbit (ISCO) radius:}

ISCO is a general relativistic prediction, and it is of enormous interest
not only for testing general relativity, but also for probing accretion processes 
and stellar evolution. This is because the accretion disc
can extend at most up to ISCO, and matter has to plunge onto the compact star beyond that.
The ISCO radius is $r_{\rm ISCO} = 6GM_{\rm G}/c^2$ for the Schwarzschild spacetime, and this 
value is somewhat different for spinning stars (see \S~\ref{Computation} for computation procedure). 
Note that for $R_{\rm e} \ge r_{\rm ISCO}$,
the disc can extend up to the stellar surface, and hence in Table~\ref{table_result}, we list the 
values of $r_{\rm orb}$, which is $r_{\rm ISCO}$ or $R_{\rm e}$, whichever is bigger 
(\S~\ref{Computation}). In this
table, we also list $r_{\rm orb} - R_{\rm e}$, which is the extent of the gap between the
accretion disc and the compact star. This gap can be very useful to understand the
accretion process and the resulting stellar evolution, as well as to interpret the X-ray
energy spectrum from accreting compact stars \citep{Bhattacharyyaetal2000}. Therefore,
$r_{\rm orb}$ and $r_{\rm orb} - R_{\rm e}$ values are important for accreting
compact stars. But, since the masses of these accreting stars have so far not been accurately
measured, the estimation of $r_{\rm orb}$ for them can suffer from large systematic errors.
In Table~\ref{table_result}, although we list the $r_{\rm orb}$ and $r_{\rm orb} - R_{\rm e}$ values 
of non-accreting ms pulsars, these will be useful even to probe accreting stars, because
these ms pulsars were spun up via accretion in LMXB phases. It is also very important to
know whether the accretion disc terminates at ISCO ($r_{\rm orb} - R_{\rm e} > 0$)
or at the stellar surface ($r_{\rm orb} - R_{\rm e} = 0$), in order to
use relativistic spectral lines observed from the disc to constrain EoS models \citep{Bhattacharyya2011}.
Table~\ref{table_result} shows that, for a given stellar mass, 
$r_{\rm orb} - R_{\rm e}$ is larger for softer EoS models, because $R_{\rm e}$ is smaller.
This table also shows that, unless the stellar mass is high, $r_{\rm orb} - R_{\rm e}$ is
usually consistent with zero, which means that the disc touches the star.

\noindent
{\bf (6) Angular momentum ($J$):}

The spin-up of a compact star to the millisecond period (see \S~\ref{Introduction}) depends on how much
$J$ it gained. The $J$ value of an ms pulsar is essentially equal to this net gain in angular momentum, 
because the initial stellar $J$ value is expected to be small as $J$ is roughly proportional to the spin 
frequency. Therefore, the $J$ values of 16 ms pulsars for eight EoS models, which we list in
Table~\ref{table_result}, also indicate the net gain in $J$ in their LMXB phase, and may provide a testbed 
to understand the torque mechanisms and the accretion processes in that phase.
This is because the net gain in $J$ depends on the spin-up and spin-down torques and
the specific process of accretion. Here we note that the $J$ value of a compact star may
increase in the accretion phase, and may decrease in the propeller phase and via electromagnetic radiation
\citep{GhoshLamb1978, Ghosh1995}, and the nature of the corresponding torques are not yet fully understood.
From Table~\ref{table_result}, we find that the $J$ values are higher for higher 
stellar mass and spin rate, as expected. They are also higher (can be by $> 50$\%) for harder EoS models (Table~\ref{table_result}).
This means, in order to attain a given mass and spin rate, a compact star with a harder EoS model may require
a significantly larger angular momentum transfer. This could, in principle, provide a way to distinguish
between EoS models using computation of the LMXB evolution.

The dimensionless angular momentum parameter ($a = cJ/GM_{\rm G}^2$) is very important to 
model the observable effects of spin on the spacetime, as well as to compare the properties of compact 
stars and black holes \citep{Bhattacharyya2011}.
Table~\ref{table_result} lists the computed $a$ values for 16 ms pulsars for eight EoS models.
Some of these compact stars, having known masses and spin rates, can be promising sources to constrain
EoS models \citep[for example, using a timing feature from the stellar surface;][]{Bogdanov2013}. Since $a$ affects 
such a timing feature, it is useful to know at least a range of $a$ values for these pulsars.
Table~\ref{table_result} provides such a range for each of 16 ms pulsars.

\noindent
{\bf (7) Moment of inertia ($I$):}

Moment of inertia of Pulsar A of the double pulsar system PSR J0737-3039 could 
be measured with up to 10\% accuracy in near future \citep{Lyneetal2004,KramerWex2009,LattimerSchutz2005}.
This will be a promising way to constrain compact star EoS models. Therefore, it may be 
interesting to check the $I$-values and their dependencies on other parameters for the pulsars 
we consider in this paper. From Table~\ref{table_result}, we find that $I$ is larger for
greater stellar masses and spin rates and for harder EoS models. This is because 
$I \sim M_{\rm G} R_{\rm e}^2$, and $R_{\rm e}$ increases with spin rate and EoS hardness.
Table~\ref{table_result} shows that $I$ is in the range of $(1.0-3.6)\times10^{45}$~g cm$^2$
for our sample of ms pulsars and EoS models. It can be useful to estimate the value of 
$A$ in $I = A M_{\rm G} R_{\rm e}^2$, especially for the computation of evolution of compact stars in the LMXB phase.
We find the average value of $A$ for our sample of ms pulsars and EoS models is $0.38\pm0.05$.
Note that this is consistent with the value for a uniform sphere.
However, the $A$-values for our strange matter EoS models are significantly higher than those
for other EoS models in our sample.


\noindent
{\bf (8) Stellar stability:}

A gravitational radiation driven nonaxisymmetric instability may set in at
a high value of the ratio of the total spinning kinetic energy to the total gravitational 
energy ($T/W$; \citet{Cooketal1994}) of compact stars. This high value is $\sim 0.08$ based on Newtonian results
\citep{Friedmanetal1986}. Table~\ref{table_result} shows that the upper limit of $T/W$ for all 16 ms pulsars
for all eight EoS models is $\approx 0.03$. This indicates that none of these pulsars is susceptible to triaxial instabilities.

\subsection{On constraining EoS models}\label{differentEoS}

The ``correct" EoS model can be narrowed down by rejecting as many theoretically proposed EoS models as possible.
For this, three independent parameters of the same compact star are to be measured (see \S~\ref{Introduction}).
In order to check if an EoS model can be rejected, one needs to compute the stable stellar configuration for 
that EoS model using two measured parameter values. Then the computed value and the measured value of the 
third parameter are to be compared. If an appropriate error bar on the measured value of the 
third parameter can be assigned, then one can 
estimate the significance with which the EoS model can be rejected. This way the EoS model rejection can be
quantitatively done by a simple comparison in one-parameter space. 
This shows why compact star structure computation using measured parameter values, as reported in this
paper, is essential. However, note that the error bar on the above mentioned
third parameter involves two errors: (1) the uncertainty in the third parameter measurement, and (2) the 
uncertainties in first and second parameter measurements, which are converted into a third parameter uncertainty.
The latter is essential to estimate the significance of EoS model rejection in one-parameter space.
In this paper, we not only demonstrate how to convert the uncertainty of a measured parameter (i.e., gravitational mass) 
into that of other parameters (see \S~\ref{Computation}), but also give the first extensive catalogue of
these parameters with uncertainties (Table~\ref{table_result}). One of these ``other" parameters (e.g., radius) of 
some of our sample compact stars could be measured in the future \citep[e.g., with {\it NICER};][]{Gendreauetal2012},
thus constraining EoS models quantitatively.

While it is useful to reject individual EoS models as mentioned above, 
it may be more important to be able to reject entire EoS classes.
This could be quantitatively done using the above procedure involving compact star structure computation, 
if three parameters for two or more
compact stars are measured. For example, the mass-radius curves for nucleonic and strange matter EoS models
have usually quite different slopes in most part of the relevant mass range (see Fig.~\ref{fig1}). 
Such different slopes imply, while the mass-radius curves for a nucleonic and a strange matter EoS models
can cross each other at one mass value, these curves can be far apart in the radius space at a sufficiently
different mass value (e.g., N1 and S2 curves in Fig.~\ref{fig1}). 
Therefore, measurement of parameters of two compact stars with sufficiently different
masses could be useful to distinguish nucleonic EoS models from strange matter EoS models.
This points towards the necessity to discover many low-mass compact stars, along with high mass stars.

\subsection{Spinning versus non-spinning configurations}\label{non-spinning}

We now briefly discuss if it was required to make the catalogue (Table~\ref{table_result}) 
computing the spinning 
configurations of compact stars, or would the easier computation of non-spinning configurations 
be sufficient? This question may arise because the spin frequency values of the ms pulsars considered here 
are in the range $\approx 132-465$ Hz; these 
are not as high as those of some of the fastest pulsars and 
are well below the mass-shed limits for the considered EoS models.
We note that the precisely measured mass values of a number of ms pulsars provide a 
tantalizing opportunity to characterize these compact stars at an unprecedented level,
and the aim of this paper is to utilize this opportunity, which requires 
calculation of spinning configurations. 

In addition, computation of spinning configurations is essential to estimate the values of some stellar 
parameters given in our catalogue, such as stellar oblateness, total and dimensionless
angular momenta and the stability indicator. The importance of these parameters have been
discussed in \S~\ref{Properties}. For example, a knowledge of the total angular momentum 
of an observed ms pulsar, which is essentially the angular momentum transferred in the LMXB phase, 
can be very useful to understand the stellar and binary evolution.

For constraining EoS models using the radius measurements of radio ms pulsars, for example with 
{\it NICER}, it is generally argued that the relatively slow spin rates (although faster than 10 ms
period) of such compact stars minimally affect the radius. However, the amount of the spin-related 
systematic error depends on what accuracy one aims to measure the equatorial radius ($R_{\rm e}$) with.
Such desired accuracy ($\xi$) is about $5$\% \citep{LattimerPrakash2001}, and {\it NICER} could also measure the stellar radius with 
a similar accuracy \citep[e.g.,][]{Gendreauetal2012}. We, therefore, compute the radius ($R$) values of
our sample of ms pulsars for all eight EoS models for non-spinning configurations (Table~\ref{table_diff}),
and calculate the percentage difference $\eta$ between $R$ and $R_{\rm e}$ using the $R_{\rm e}$-values
for spinning configurations (Table~\ref{table_result}).
If the EoS models are constrained using a $R_{\rm e}$-value measured with a percentage accuracy of $\pm \xi$ and using 
theoretical non-spinning configurations, then there may be $(\eta/2\xi)\times100$\% of systematic error or bias in 
the allowed EoS models (Table~\ref{table_diff}). This means, $\sim (\eta/2\xi)\times100$\% of the allowed EoS models
would be falsely allowed because of the systematic difference between $R$ and $R_{\rm e}$, while a similar 
number of EoS models would be falsely ruled out.
This bias is quite high for the faster spinning pulsars of our sample for our EoS models (e.g., $19-47$\% for 
PSR J1903+0327; Table~\ref{table_diff}). Note that even though the bias is smaller for slower pulsars,
estimation of such a bias, as given in Table~\ref{table_diff}, will be required to know the reliability of constraints on EoS models.
Moreover, for a given $\eta$ value, $\xi$ is expected to decrease with the availability of
better instruments in the future. This will increase the bias values given in Table~\ref{table_diff},
which shows the usefulness of the first extensive tabulation of these values and the computation of spinning 
configurations in this paper.

\section{Summary and conclusions}\label{Summary}

In this paper, we report a catalogue of the computed parameter values of
a number of observed ms pulsars. This, to the best of our knowledge,
is the first such catalogue of this kind and extent, and gives the best 
characterization of observed ms pulsars so far. 
Note that the EoS models used to make this catalogue are chosen from different classes,
and they can support the maximum observed compact star mass. Such a catalogue could be
made, because recent precise mass measurements of a number of ms pulsars
has created a sizable population of compact stars with two precisely
known parameters. From these two parameters, and an assumed EoS model,
we compute other parameter values of ms pulsars. We also estimate uncertainties
of these parameters from the errors in measured stellar masses.
These parameters, which are calculated for different EoS models, can be
useful to constrain these models. Furthermore, the catalogue may provide
a testbed to probe the physical processes governing the compact star evolution.

We also discuss how an individual EoS model can be quantitatively rejected in 
one-parameter space by converting the uncertainty of a measured parameter into
an uncertainty of another parameter for that EoS model.
In addition to constraining individual EoS models, it is important to constrain EoS classes.
Here we suggest that radius measurements with shorter observations of at least two ms pulsars, 
ideally one of low mass and another of high mass, could be
more useful to discriminate between nucleonic and strange matter EoS models than the
radius measurement with observation of one ms pulsar utilizing the entire observation time.
Therefore, while high mass compact stars are usually searched for to constrain EoS models,
finding a population of low mass ms pulsars can also be rewarding.

For our sample of ms pulsars and diverse EoS models, 
we find that the largest value of the central density is $\rho_{\rm c} \sim 5.8 \rho_{\rm sat}$ 
in the case of PSR~J1614-2230, i.e., the ms pulsar in our sample with the largest gravitational mass.
This is much less than the expected maximum value $13 \rho_{\rm sat}$ 
\citep[see][]{LattimerPrakash2005,LattimerPrakash2010}.
We argue that the $\rho_{\rm c}$-values of at most a small fraction of compact stars could be much larger
than the largest value found in this paper.
The lower density values favored by our computations can have implications for stellar 
formation and evolution, for the constituents of stellar cores, and for mapping of observables 
to the parametrized EoS models in an attempt to constrain these models
\citep{Raitheletal2016}. 

An important point of this paper is the computation of neutron star properties such as oblateness, angular
momentum and the stability indicator that cannot be obtained from computations of non-spinning configurations.
We also compute the bias values in the allowed EoS models,
if the EoS models are constrained using a stellar equatorial radius
measured with $\pm 5$\% accuracy and using theoretical non-spinning configurations.
This, to the best of our knowledge, is heretofore the most detailed study of this
bias, and will be useful to reliably constrain EoS models. We find that this bias
could be significant for certain combinations of parameter values.
But even when it is smaller, it should be estimated to know the reliability of constraints
on EoS models. Such estimation requires computation of stellar spinning configurations.

\section*{Acknowledgements}

AVT acknowledges funding from SERB-DST (File number EMR/2016/002033).
This work has been partially supported by the COST Action MP1304
Exploring fundamental physics with compact stars (NewCompstar).



\begin{table*}
\centering
\caption{List of ms pulsars with measured gravitational mass and less than 10 ms spin-period (see \S~\ref{EoS}).}
\begin{tabular}{rlcll}
\hline
No. & Pulsar & Spin-period [frequency] & Mass & References\footnotemark[1] \\
 & name & (ms [Hz]) & $(M_{\odot})$ & \\
\hline
1 & J1903+0327 & 2.15 [465.1] & $1.667_{-0.021}^{+0.021}$ & 1, 2 \\
2 & J2043+1711 & 2.40 [416.7] & $1.41_{-0.18}^{+0.21}$ & 3, 4, 5 \\
3 & J0337+1715 & 2.73 [366.0] & $1.4378_{-0.0013}^{+0.0013}$ & 6 \\
4 & J1909-3744 & 2.95 [339.0] & $1.540_{-0.027}^{+0.027}$ & 7, 8 \\
5 & J1614-2230 & 3.15 [317.5] & $1.928_{-0.017}^{+0.017}$ & 9, 5 \\
6 & J1946+3417 & 3.17 [315.5] & $1.832_{-0.028}^{+0.028}$ & 10, 11 \\
7 & J1911-5958A & 3.27 [305.8] & $1.33_{-0.11}^{+0.11}$ & 12, 13 \\
8 & J0751+1807 & 3.48 [287.4] & $1.64_{-0.15}^{+0.15}$ & 14, 8 \\
9 & J2234+0611 & 3.58 [279.3] & $1.393_{-0.013}^{+0.013}$ & 15, 11 \\
10 & J1807-2500B & 4.19 [238.7] & $1.3655_{-0.0021}^{+0.0021}$ & 16 \\
11 & J1713+0747 & 4.57 [218.9] & $1.33_{-0.08}^{+0.09}$ & 17, 8 \\
12 & J1012+5307 & 5.26 [190.1] & $1.83_{-0.11}^{+0.11}$ & 18, 11 \\
13 & B1855+09 & 5.36 [186.6] & $1.30_{-0.10}^{+0.11}$ & 19, 5 \\
14 & J0437-4715 & 5.76 [173.6] & $1.44_{-0.07}^{+0.07}$ & 20, 21 \\
15 & J1738+0333 & 5.85 [170.9] & $1.47_{-0.06}^{+0.07}$ & 22, 23 \\
16 & J1918-0642 & 7.60 [131.6] & $1.18_{-0.09}^{+0.10}$ & 24, 5 \\
\hline
\end{tabular}
\begin{flushleft}
$^1$[1] \citet{Championetal2008}; [2] \citet{Freireetal2011}; [3] \citet{Abdoetal2010}; [4] \citet{Guillemotetal2012}; [5] \citet{Fonsecaetal2016}; [6] \citet{Ransometal2014}; [7] \citet{Jacobyetal2003}; [8] \citet{Desvignesetal2016}; [9] \citet{Demorestetal2010}; [10] \citet{Barretal2013}; [11] \citet{OzelFreire2016}; [12]\citet{DAmicoetal2001}; [13] \citet{Corongiuetal2012}; [14] \citet{Lundgrenetal1993}; [15] \citet{Denevaetal2013}; [16] \citet{Lynchetal2012}; [17] \citet{Fosteretal1993}; [18] \citet{Nicastroetal1995}; [19] \citet{Segelsteinetal1986}; [20] \citet{Johnstonetal1993}; [21] \citet{Reardonetal2016}; [22] \citet{Jacobyetal2007}; [23] \citet{Antoniadisetal2012}; [24] \citet{EdwardsBailes2001}.
\end{flushleft}
\label{table_pulsar}
\end{table*}

\begin{table*}
\centering
\caption{List of used EoS models and the maximum gravitational mass each model can support 
in non-spinning configuration (see \S~\ref{EoS}).}
\begin{tabular}{llllll}
\hline
No. & EoS & Type & Brief & Maximum & References\footnotemark[1] \\
    & model & of EoS & description & non-spinning & \\
    & & & & mass ($M_\odot$) & \\
\hline
1 & N1 & Nucleonic & A18+$\delta v$+UIX & 2.199 & 1 \\
2 & N2 & Nucleonic & Chiral sigma model & 2.589 & 2 \\ 
3 & N3 & Nucleonic & Relativistic mean field (RMF) model & 2.282 & 3, 4\\
4 & Hp1 & Hyperonic & RMF model: n,p,$\Lambda$          & 2.099 & 5 \\
5 & Hp2 & Hyperonic & RMF model: n,p,$\Lambda$,$\Sigma^-$,$\Sigma^0$,$\Sigma^+$,$\Xi^-$,$\Xi^0$ & 1.976 & 4\\
6 & S1 & Strange matter & $B_{\rm eff}^{1/4} = 138$ MeV, $a_4 = 0.8$ & 2.093 & 6, 7, 8, 9 \\
7 & S2 & Strange matter & $B_{\rm eff}^{1/4} = 125$ MeV, $a_4 = 0.5$ & 2.479 & 6, 7, 8, 9 \\
8 & Hb1 & Hybrid & A18+$\delta v$+UIX and $B_{\rm eff}^{1/4} = 138$ MeV, $a_4 = 0.4$ & 2.103 & 1, 9, 10, 11 \\
\hline
\end{tabular}
\begin{flushleft}
$^1$[1] \citet{Akmaletal1998}; [2] \citet{Sahuetal1993}; [3] \citet{SugaharaToki1994};
[4] \citet{providencia2013}; [5] \citet{Baniketal2014};
[6] \citet{Fra01}; [7] \citet{Alf05}; [8] \citet{weis11}; [9] \citet{Bhattacharyyaetal2016};
[10] \citet{glendenning1992}; [11] \citet{glendenning1996}.
\end{flushleft}
\label{table_EoS}
\end{table*}

\begin{table*}
\centering
\caption{Theoretically computed parameter values of pulsars (Table~\ref{table_pulsar}) for different EoS models 
(\S~\ref{EoS} and \ref{Results}).}
\scalebox{0.80}{%
\begin{tabular}{l@{}c@{\hspace{2mm}}c@{\hspace{2mm}}c@{\hspace{2mm}}c@{\hspace{2mm}}c@{\hspace{2mm}}c@{\hspace{2mm}}c@{\hspace{2mm}}c@{\hspace{2mm}}c@{\hspace{2mm}}c@{\hspace{2mm}}c@{\hspace{2mm}}c}
\hline
  EoS\footnotemark[1] & $\rho_{\rm c}$\footnotemark[2] & $M_{\rm 0}$\footnotemark[3] & $R_{\rm e}$\footnotemark[4] & $R_{\rm e}/r_{\rm g}$\footnotemark[5] & $R_{\rm p}$\footnotemark[6] & $R_{\rm p}/R_{\rm e}$ & $r_{\rm orb}$\footnotemark[7] & $r_{\rm orb}-R_{\rm e}$ & $J$\footnotemark[8] & $cJ/GM_{\rm G}^2$\footnotemark[9] & $I$\footnotemark[10] & $T/W$\footnotemark[11]\\
 & ($10^{14}$ & ($M_{\odot}$) & (km) & & (km) & & (km) & (km) & ($10^{49}$ & & ($10^{45}$ & \\
 & g cm$^{-3}$) & & & & & & & & g cm$^2$ s$^{-1}$) & & g cm$^2$) & \\
\hline
\multicolumn{13}{c}{(1) PSR J1903+0327: $M_{\rm G}$\footnotemark[12]$ = 1.667_{-0.021}^{+0.021} M_\odot$, $\nu$\footnotemark[13]$ = 465.1$~Hz}\\
\hline
N1 & $11.58_{-0.17}^{+0.18}$ & $1.89_{-0.03}^{+0.03}$ & $11.64_{-0.02}^{+0.02}$ & $4.73_{-0.07}^{+0.07}$ & $11.22_{-0.01}^{+0.01}$ & $0.96_{-0.00}^{+0.00}$ & $13.40_{-0.16}^{+0.16}$ & $1.76_{-0.18}^{+0.18}$ & $0.51_{-0.01}^{+0.01}$ & $0.21_{-0.00}^{+0.00}$ & $1.74_{-0.03}^{+0.03}$ & $0.01_{-0.00}^{+0.00}$\\
N2 & $4.35_{-0.04}^{+0.04}$ & $1.83_{-0.03}^{+0.03}$ & $15.73_{-0.01}^{+0.01}$ & $6.39_{-0.08}^{+0.08}$ & $14.24_{-0.02}^{+0.02}$ & $0.89_{-0.00}^{+0.00}$ & $15.73_{-0.01}^{+0.01}$ & $0.00_{-0.00}^{+0.00}$ & $0.87_{-0.02}^{+0.02}$ & $0.36_{-0.00}^{+0.00}$ & $2.99_{-0.05}^{+0.06}$ & $0.03_{-0.00}^{+0.00}$\\
N3 & $6.28_{-0.09}^{+0.10}$ & $1.84_{-0.03}^{+0.03}$ & $14.95_{-0.03}^{+0.03}$ & $6.07_{-0.09}^{+0.09}$ & $13.80_{-0.00}^{+0.00}$ & $0.91_{-0.00}^{+0.00}$ & $14.95_{-0.03}^{+0.03}$ & $0.00_{-0.00}^{+0.00}$ & $0.75_{-0.01}^{+0.01}$ & $0.31_{-0.00}^{+0.00}$ & $2.57_{-0.04}^{+0.04}$ & $0.02_{-0.00}^{+0.00}$\\
Hp1 & $7.59_{-0.15}^{+0.16}$ & $1.86_{-0.03}^{+0.03}$ & $13.65_{-0.02}^{+0.02}$ & $5.54_{-0.08}^{+0.08}$ & $12.83_{-0.00}^{+0.00}$ & $0.93_{-0.00}^{+0.00}$ & $13.65_{-0.02}^{+0.02}$ & $0.00_{-0.00}^{+0.00}$ & $0.67_{-0.01}^{+0.01}$ & $0.27_{-0.00}^{+0.00}$ & $2.28_{-0.04}^{+0.04}$ & $0.02_{-0.00}^{+0.00}$\\
Hp2 & $6.66_{-0.17}^{+0.18}$ & $1.84_{-0.03}^{+0.03}$ & $14.92_{-0.04}^{+0.04}$ & $6.06_{-0.09}^{+0.09}$ & $13.79_{-0.01}^{+0.01}$ & $0.91_{-0.00}^{+0.00}$ & $14.92_{-0.04}^{+0.04}$ & $0.00_{-0.00}^{+0.00}$ & $0.75_{-0.01}^{+0.01}$ & $0.31_{-0.00}^{+0.00}$ & $2.56_{-0.03}^{+0.03}$ & $0.02_{-0.00}^{+0.00}$\\
S1 & $6.91_{-0.10}^{+0.11}$ & $2.09_{-0.03}^{+0.03}$ & $12.15_{-0.03}^{+0.02}$ & $4.93_{-0.05}^{+0.05}$ & $11.52_{-0.03}^{+0.03}$ & $0.94_{-0.00}^{+0.00}$ & $13.22_{- 0.16}^{+0.15}$ & $1.08_{-0.13}^{+0.13}$ & $0.66_{-0.01}^{+0.01}$ & $0.27_{-0.00}^{+0.00}$ & $2.27_{-0.04}^{+0.04}$ & $0.02_{-0.00}^{+0.00}$\\
S2 & $4.09_{-0.03}^{+0.04}$ & $1.95_{-0.03}^{+0.03}$ & $14.12_{-0.04}^{+0.04}$ & $5.74_{-0.06}^{+0.05}$ & $12.88_{-0.04}^{+0.04}$ & $0.90_{-0.00}^{+0.00}$ & $14.12_{-0.04}^{+0.04}$ & $0.00_{-0.00}^{+0.00}$ & $0.87_{-0.02}^{+0.02}$ & $0.36_{-0.00}^{+0.00}$ & $2.98_{-0.06}^{+0.06}$ & $0.03_{-0.00}^{+0.00}$\\
Hb1 & $11.74_{-0.18}^{+0.18}$ & $1.89_{-0.03}^{+0.03}$ & $11.52_{-0.02}^{+0.02}$ & $4.68_{-0.06}^{+0.07}$ & $11.12_{-0.01}^{+0.01}$ & $0.96_{-0.00}^{+0.00}$ & $13.41_{-0.17}^{+0.16}$ & $1.90_{-0.18}^{+0.18}$ & $0.50_{-0.01}^{+0.01}$ & $0.21_{-0.00}^{+0.00}$ & $1.73_{-0.03}^{+0.03}$ & $0.01_{-0.00}^{+0.00}$\\
\hline
\multicolumn{13}{c}{(2) PSR J2043+1711: $M_{\rm G} = 1.41_{-0.18}^{+0.21} M_\odot$, $\nu = 416.7$~Hz}\\
\hline
N1 & $9.80_{-1.07}^{+1.46}$ & $1.56_{-0.22}^{+0.27}$ & $11.80_{-0.17}^{+0.12}$ & $5.67_{-0.80}^{+0.90}$ & $11.37_{-0.08}^{+0.04}$ & $0.96_{-0.01}^{+0.01}$ & $11.80_{-0.00}^{+1.34}$ & $0.00_{-0.00}^{+1.51}$ & $0.36_{-0.06}^{+0.07}$ & $0.21_{-0.02}^{+0.02}$ & $1.39_{-0.23}^{+0.28}$ & $0.01_{-0.00}^{+0.00}$\\
N2 & $3.96_{-0.26}^{+0.35}$ & $1.53_{-0.21}^{+0.25}$ & $15.42_{-0.12}^{+0.12}$ & $7.41_{-0.91}^{+1.02}$ & $14.09_{-0.26}^{+0.25}$ & $0.91_{-0.01}^{+0.01}$ & $15.42_{-0.12}^{+0.12}$ & $0.00_{-0.00}^{+0.00}$ & $0.60_{-0.11}^{+0.14}$ & $0.34_{-0.02}^{+0.02}$ & $2.30_{-0.42}^{0.52}$ & $0.03_{-0.00}^{+0.00}$\\
N3 & $5.32_{-0.61}^{+0.82}$ & $1.53_{-0.21}^{+0.25}$ & $15.09_{-0.24}^{+0.21}$ & $7.24_{-1.03}^{+1.17}$ & $13.93_{-0.02}^{+0.00}$ & $0.91_{-0.01}^{+0.01}$ & $15.09_{-0.24}^{+0.21}$ & $0.00_{-0.00}^{+0.00}$ & $0.55_{-0.08}^{+0.10}$ & $0.31_{-0.03}^{+0.03}$ & $2.08_{-0.32}^{+0.37}$ & $0.02_{-0.00}^{+0.00}$\\
Hp1 &  $6.20_{-0.64}^{+1.14}$ & $1.54_{-0.21}^{+0.26}$ & $13.65_{-0.07}^{+0.01}$ & $6.55_{-0.88}^{+0.97}$ & $12.86_{-0.11}^{+0.05}$ & $0.94_{-0.01}^{+0.01}$ & $13.65_{-0.07}^{+0.01}$ & $0.00_{-0.00}^{+0.00}$ & $0.47_{-0.08}^{+0.10}$ & $0.27_{-0.02}^{+0.02}$ & $1.80_{-0.32}^{+0.37}$ & $0.02_{-0.00}^{+0.00}$\\
Hp2 & $5.32_{-0.61}^{+1.09}$ & $1.53_{-0.21}^{+0.25}$ & $15.09_{-0.25}^{+0.21}$ & $7.24_{-1.04}^{+1.18}$ & $13.93_{-0.02}^{+0.00}$ & $0.91_{-0.01}^{+0.01}$ & $15.09_{-0.25}^{+0.21}$ & $0.00_{-0.00}^{+0.00}$ & $0.55_{-0.08}^{+0.10}$ & $0.31_{-0.03}^{+0.03}$ & $2.08_{-0.32}^{+0.36}$ & $0.02_{-0.00}^{+0.00}$\\
S1 & $5.95_{-0.49}^{+0.79}$ & $1.73_{-0.24}^{+0.29}$ & $11.69_{-0.38}^{+0.34}$ & $5.61_{-0.58}^{+0.61}$ & $11.14_{-0.42}^{+0.37}$ & $0.95_{-0.00}^{+0.00}$ & $11.69_{-0.38}^{+1.19}$ & $0.00_{-0.00}^{+0.85}$ & $0.46_{-0.09}^{+0.10}$ & $0.26_{-0.02}^{+0.02}$ & $1.75_{-0.34}^{+0.41}$ & $0.02_{-0.00}^{+0.00}$\\
S2 & $3.73_{-0.22}^{+0.31}$ & $1.63_{-0.22}^{+0.27}$ & $13.46_{-0.48}^{+0.47}$ & $6.46_{-0.65}^{+0.68}$ & $12.43_{-0.49}^{+0.52}$ & $0.91_{-0.00}^{+0.01}$ & $13.46_{-0.48}^{+0.47}$ & $0.00_{-0.00}^{+0.00}$ & $0.59_{-0.12}^{-0.15}$ & $0.34_{-0.02}^{+0.02}$ & $2.26_{-0.44}^{+0.56}$ & $0.03_{-0.00}^{+0.00}$\\
Hb1 & $9.87_{-1.08}^{+1.51}$ & $1.56_{-0.22}^{+0.27}$ & $11.65_{-0.13}^{+0.08}$ & $5.59_{-0.78}^{+0.86}$ & $11.23_{-0.06}^{+0.00}$ & $0.96_{-0.01}^{+0.01}$ & $11.65_{-0.00}^{+1.50}$ & $0.00_{-0.00}^{+1.64}$ & $0.36_{-0.06}^{+0.07}$ & $0.21_{-0.02}^{+0.02}$ & $1.37_{-0.24}^{+0.28}$ & $0.01_{-0.00}^{+0.00}$\\
\hline
\multicolumn{13}{c}{(3) PSR J0337+1715: $M_{\rm G} = 1.4378_{-0.0013}^{+0.0013} M_\odot$, $\nu = 366.0$~Hz}\\
\hline
N1 & $10.01_{-0.01}^{+0.01}$ & $1.59_{-0.00}^{+0.00}$ & $11.73_{-0.00}^{+0.00}$ & $5.52_{-0.01}^{+0.01}$ & $11.40_{-0.00}^{+0.00}$ & $0.97_{-0.00}^{+0.00}$ & $11.82_{-0.01}^{+0.01}$ & $0.09_{-0.01}^{+0.01}$ & $0.32_{-0.00}^{+0.00}$ & $0.18_{-0.00}^{+0.00}$ & $1.42_{-0.00}^{+0.00}$ & $0.01_{-0.00}^{+0.00}$\\
N2 & $4.03_{-0.00}^{+0.00}$ & $1.56_{-0.00}^{+0.00}$ & $15.27_{-0.00}^{+0.00}$ & $7.19_{-0.01}^{+0.01}$ & $14.28_{-0.00}^{+0.00}$ & $0.93_{-0.00}^{+0.00}$ & $15.27_{-0.00}^{+0.00}$ & $0.00_{-0.00}^{+0.00}$ & $0.54_{-0.00}^{+0.00}$ & $0.29_{-0.00}^{+0.00}$ & $2.33_{-0.00}^{+0.00}$ & $0.02_{-0.00}^{+0.00}$\\
N3 & $5.47_{-0.00}^{+0.00}$ & $1.56_{-0.00}^{+0.00}$ & $14.89_{-0.00}^{+0.00}$ & $7.01_{-0.01}^{+0.01}$ & $14.04_{-0.00}^{+0.00}$ & $0.94_{-0.00}^{+0.00}$ & $14.89_{-0.00}^{+0.00}$ & $0.00_{-0.00}^{+0.00}$ & $0.48_{-0.00}^{+0.00}$ & $0.27_{-0.00}^{+0.00}$ & $2.10_{-0.00}^{+0.00}$ & $0.02_{-0.00}^{+0.00}$\\
Hp1 &  $6.38_{-0.01}^{+0.01}$ & $1.58_{-0.00}^{+0.00}$ & $13.55_{-0.00}^{+0.00}$ & $6.38_{-0.01}^{+0.01}$ & $12.96_{-0.00}^{+0.00}$ & $0.95_{-0.00}^{+0.00}$ & $13.55_{-0.00}^{+0.00}$ & $0.00_{-0.00}^{+0.00}$ & $0.42_{-0.00}^{+0.00}$ & $0.23_{-0.00}^{+0.00}$ & $1.84_{-0.00}^{+0.00}$ & $0.01_{-0.00}^{+0.00}$\\
Hp2 & $5.47_{-0.00}^{+0.00}$ & $1.56_{-0.00}^{+0.00}$ & $14.89_{-0.00}^{+0.00}$ & $7.01_{-0.01}^{+0.01}$ & $14.04_{-0.00}^{+0.00}$ & $0.94_{-0.00}^{+0.00}$ & $14.89_{-0.00}^{+0.00}$ & $0.00_{-0.00}^{+0.00}$ & $0.48_{-0.00}^{+0.00}$ & $0.27_{-0.00}^{+0.00}$ & $2.10_{-0.00}^{+0.00}$ & $0.02_{-0.00}^{+0.00}$\\
S1 & $6.07_{-0.00}^{+0.00}$ & $1.77_{-0.00}^{+0.00}$ & $11.68_{-0.00}^{+0.00}$ & $5.50_{-0.00}^{+0.00}$ & $11.30_{-0.00}^{+0.00}$ & $0.96_{-0.00}^{+0.00}$ & $11.68_{-0.00}^{+0.00}$ & $0.00_{-0.00}^{+0.00}$ & $0.41_{-0.00}^{+0.00}$ & $0.23_{-0.00}^{+0.00}$ & $1.78_{-0.00}^{+0.00}$ & $0.01_{-0.00}^{+0.00}$\\
S2 & $3.80_{-0.00}^{+0.00}$ & $1.67_{-0.00}^{+0.00}$ & $13.43_{-0.00}^{+0.00}$ & $6.32_{-0.00}^{+0.00}$ & $12.68_{-0.00}^{+0.00}$ & $0.94_{-0.00}^{+0.00}$ & $13.43_{-0.00}^{+0.00}$ & $0.00_{-0.00}^{+0.00}$ & $0.53_{-0.00}^{+0.00}$ & $0.29_{-0.00}^{+0.00}$ & $2.30_{-0.00}^{+0.00}$ & $0.02_{-0.00}^{+0.00}$\\
Hb1 & $10.10_{-0.01}^{+0.01}$ & $1.60_{-0.00}^{+0.00}$ & $11.58_{-0.00}^{+0.00}$ & $5.45_{-0.01}^{+0.01}$ & $11.27_{-0.00}^{+0.00}$ & $0.97_{-0.00}^{+0.00}$ & $11.82_{-0.01}^{+0.01}$ & $0.24_{-0.01}^{+0.01}$ & $0.32_{-0.00}^{+0.00}$ & $0.18_{-0.00}^{+0.00}$ & $1.40_{-0.00}^{+0.00}$ & $0.01_{-0.00}^{+0.00}$\\
\hline
\multicolumn{13}{c}{(4) PSR J1909-3744: $M_{\rm G} = 1.540_{-0.027}^{+0.027} M_\odot$, $\nu = 339.0$~Hz}\\
\hline
N1 & $10.71_{-0.18}^{+0.20}$ & $1.72_{-0.03}^{+0.04}$ & $11.63_{-0.02}^{+0.02}$ & $5.12_{-0.10}^{+0.10}$ & $11.38_{-0.01}^{+0.01}$ & $0.97_{-0.00}^{+0.00}$ & $12.69_{-0.21}^{+0.21}$ & $1.05_{-0.22}^{+0.24}$ & $0.33_{-0.01}^{+0.01}$ & $0.16_{-0.00}^{+0.00}$ & $1.55_{-0.04}^{+0.04}$ & $0.01_{-0.00}^{+0.00}$\\
N2 & $4.22_{-0.04}^{+0.05}$ & $1.68_{-0.03}^{+0.03}$ & $15.27_{-0.02}^{+0.02}$ & $6.71_{-0.11}^{+0.11}$ & $14.47_{-0.03}^{+0.03}$ & $0.94_{-0.00}^{+0.00}$ & $15.27_{-0.02}^{+0.02}$ & $0.00_{-0.00}^{+0.00}$ & $0.55_{-0.01}^{+0.01}$ & $0.26_{-0.00}^{+0.00}$ & $2.56_{-0.06}^{+0.07}$ & $0.02_{-0.00}^{+0.00}$\\
N3 & $5.90_{-0.11}^{+0.11}$ & $1.69_{-0.03}^{+0.03}$ & $14.72_{-0.02}^{+0.02}$ & $6.47_{-0.12}^{+0.13}$ & $14.08_{-0.01}^{+0.00}$ & $0.95_{-0.00}^{+0.00}$ & $14.72_{-0.02}^{+0.02}$ & $0.00_{-0.00}^{+0.00}$ & $0.48_{-0.01}^{+0.01}$ & $0.23_{-0.00}^{+0.00}$ & $2.26_{-0.05}^{+0.05}$ & $0.01_{-0.00}^{+0.00}$\\
Hp1 & $6.94_{-0.15}^{+0.16}$ & $1.70_{-0.03}^{+0.03}$ & $13.49_{-0.01}^{+0.01}$ & $5.93_{-0.11}^{+0.11}$ & $13.01_{-0.00}^{+0.00}$ & $0.96_{-0.00}^{+0.00}$ & $13.49_{-0.01}^{+0.01}$ & $0.00_{-0.00}^{+0.00}$ & $0.43_{-0.01}^{+0.01}$ & $0.21_{-0.00}^{+0.00}$ & $2.01_{-0.05}^{+0.05}$ & $0.01_{-0.00}^{+0.00}$\\
Hp2 & $6.01_{-0.16}^{+0.17}$ & $1.68_{-0.03}^{+0.03}$ & $14.72_{-0.03}^{+0.03}$ & $6.47_{-0.12}^{+0.13}$ & $14.08_{-0.01}^{+0.01}$ & $0.95_{-0.00}^{+0.00}$ & $14.72_{-0.03}^{+0.03}$ & $0.00_{-0.00}^{+0.00}$ & $0.48_{-0.01}^{+0.01}$ & $0.23_{-0.00}^{+0.00}$ & $2.26_{-0.05}^{+0.05}$ & $0.01_{-0.00}^{+0.00}$\\
S1 & $6.47_{-0.11}^{+0.11}$ & $1.91_{-0.04}^{+0.04}$ & $11.84_{-0.04}^{+0.04}$ & $5.20_{-0.07}^{+0.07}$ & $11.51_{-0.05}^{+0.04}$ & $0.97_{-0.00}^{+0.00}$ & $12.44_{-0.21}^{+0.21}$ & $0.60_{-0.17}^{+0.17}$ & $0.42_{-0.01}^{+0.01}$ & $0.20_{-0.00}^{+0.00}$ & $1.98_{-0.05}^{+0.05}$ & $0.01_{-0.00}^{+0.00}$\\
S2 & $3.96_{-0.04}^{+0.04}$ & $1.80_{-0.03}^{+0.04}$ & $13.62_{-0.06}^{+0.06}$ & $5.99_{-0.08}^{+0.08}$ & $12.99_{-0.06}^{+0.06}$ & $0.95_{-0.00}^{+0.00}$ & $13.62_{-0.06}^{+0.06}$ & $0.00_{-0.00}^{+0.00}$ & $0.54_{-0.02}^{+0.02}$ & $0.26_{-0.00}^{+0.00}$ & $2.55_{-0.07}^{+0.07}$ & $0.02_{-0.00}^{+0.00}$\\
Hb1 & $10.83_{-0.20}^{+0.20}$ & $1.73_{-0.04}^{+0.03}$ & $11.50_{-0.02}^{+0.02}$ & $5.05_{-0.09}^{+0.10}$ & $11.26_{-0.01}^{+0.01}$ & $0.98_{-0.00}^{+0.00}$ & $12.69_{-0.22}^{+0.21}$ & $1.19_{-0.24}^{+0.23}$ & $0.33_{-0.01}^{+0.01}$ & $0.16_{-0.00}^{+0.00}$ & $1.54_{-0.04}^{+0.04}$ & $0.01_{-0.00}^{+0.00}$\\
\hline
\end{tabular}}
\begin{flushleft}
$^1$Equation of state models (Table~\ref{table_EoS} and Fig.~\ref{fig1}).\\
$^2$Central density.\\
$^3$Rest mass.\\
$^4$Equatorial radius.\\
$^5$Inverse of stellar compactness. Here, $r_{\rm g}$ is the Schwarzschild radius.\\
$^6$Polar radius.\\
$^7$Radius of the innermost stable circular orbit, or the stellar equatorial radius, whichever is bigger.\\
$^8$Total angular momentum.\\
$^9$Dimensionless angular momentum parameter; $c$ is the speed of light in vacuum, and $G$ is the 
	gravitational constant.\\
$^{10}$Moment of inertia.\\
$^{11}$Ratio of the total spinning kinetic energy to the total gravitational energy.\\
$^{12}$Gravitational mass.\\
$^{13}$Spin frequency.\\
Table 3 is continued on the next page.
\end{flushleft}
\label{table_result}
\end{table*}

\begin{table*}
\centering
{\bf Table~\ref{table_result}} (continued)\\
\scalebox{0.80}{%
\begin{tabular}{l@{}c@{\hspace{2mm}}c@{\hspace{2mm}}c@{\hspace{2mm}}c@{\hspace{2mm}}c@{\hspace{2mm}}c@{\hspace{2mm}}c@{\hspace{2mm}}c@{\hspace{2mm}}c@{\hspace{2mm}}c@{\hspace{2mm}}c@{\hspace{2mm}}c}
\hline
  EoS\footnotemark[1] & $\rho_{\rm c}$\footnotemark[2] & $M_{\rm 0}$\footnotemark[3] & $R_{\rm e}$\footnotemark[4] & $R_{\rm e}/r_{\rm g}$\footnotemark[5] & $R_{\rm p}$\footnotemark[6] & $R_{\rm p}/R_{\rm e}$ & $r_{\rm orb}$\footnotemark[7] & $r_{\rm orb}-R_{\rm e}$ & $J$\footnotemark[8] & $cJ/GM_{\rm G}^2$\footnotemark[9] & $I$\footnotemark[10] & $T/W$\footnotemark[11]\\
 & ($10^{14}$ & ($M_{\odot}$) & (km) & & (km) & & (km) & (km) & ($10^{49}$ & & ($10^{45}$ & \\
 & g cm$^{-3}$) & & & & & & & & g cm$^2$ s$^{-1}$) & & g cm$^2$) & \\
\hline
\multicolumn{13}{c}{(5) PSR J1614-2230: $M_{\rm G} = 1.928_{-0.017}^{+0.017} M_\odot$, $\nu = 317.5$~Hz}\\
\hline
N1 & $14.58_{-0.25}^{+0.25}$ & $2.25_{-0.03}^{+0.02}$ & $11.19_{-0.03}^{+0.03}$ & $3.93_{-0.04}^{+0.05}$ & $11.05_{-0.03}^{+0.03}$ & $0.98_{-0.00}^{+0.00}$ & $15.99_{-0.15}^{+0.14}$ & $4.79_{-0.18}^{+0.17}$ & $0.41_{-0.00}^{+0.00}$ & $0.13_{-0.00}^{+0.00}$ & $2.06_{-0.02}^{+0.02}$ & $0.00_{-0.00}^{+0.00}$\\
N2 & $5.06_{-0.05}^{+0.04}$ & $2.16_{-0.02}^{+0.02}$ & $15.43_{-0.00}^{+0.00}$ & $5.42_{-0.04}^{+0.05}$ & $14.83_{-0.01}^{+0.01}$ & $0.96_{-0.00}^{+0.00}$ & $15.90_{-0.19}^{+0.38}$ & $0.47_{-0.19}^{+0.37}$ & $0.71_{-0.01}^{+0.01}$ & $0.22_{-0.00}^{+0.00}$ & $3.57_{-0.05}^{+0.04}$ & $0.01_{-0.00}^{+0.00}$\\
N3 & $7.94_{-0.11}^{+0.13}$ & $2.17_{-0.02}^{+0.02}$ & $14.25_{-0.03}^{+0.03}$ & $5.00_{-0.05}^{+0.05}$ & $13.84_{-0.02}^{+0.02}$ & $0.97_{-0.00}^{+0.00}$ & $15.76_{-0.12}^{+0.13}$ & $1.51_{-0.14}^{+0.16}$ & $0.59_{-0.01}^{+0.01}$ & $0.18_{-0.00}^{+0.00}$ & $2.95_{-0.03}^{+0.03}$ & $0.01_{-0.00}^{+0.00}$\\
Hp1 & $10.70_{-0.28}^{+0.31}$ & $2.20_{-0.02}^{+0.02}$ & $13.01_{-0.05}^{+0.04}$ & $4.57_{-0.06}^{+0.06}$ & $12.74_{-0.04}^{+0.03}$ & $0.98_{-0.00}^{+0.00}$ & $15.80_{-0.14}^{+0.15}$ & $2.79_{-0.19}^{+0.19}$ & $0.51_{-0.00}^{+0.00}$ & $0.16_{-0.00}^{+0.00}$ & $2.58_{-0.02}^{+0.02}$ & $0.01_{-0.00}^{+0.00}$\\
Hp2 & $11.46_{-0.58}^{+0.71}$ & $2.18_{-0.02}^{+0.02}$ & $13.66_{-0.13}^{+0.11}$ & $4.80_{-0.09}^{+0.08}$ & $13.34_{-0.11}^{+0.09}$ & $0.97_{-0.00}^{+0.00}$ & $15.78_{-0.13}^{+0.14}$ & $2.12_{-0.25}^{+0.27}$ & $0.54_{-0.00}^{+0.00}$ & $0.16_{-0.00}^{+0.00}$ & $2.69_{-0.02}^{+0.01}$ & $0.01_{-0.00}^{+0.00}$\\
S1 & $9.22_{-0.22}^{+0.23}$ & $2.47_{-0.03}^{+0.03}$ & $12.17_{-0.00}^{+0.00}$ & $4.27_{-0.04}^{+0.04}$ & $11.95_{-0.00}^{+0.00}$ & $0.98_{-0.00}^{+0.00}$ & $15.67_{-0.15}^{+0.14}$ & $3.50_{-0.14}^{+0.15}$ & $0.54_{-0.01}^{+0.01}$ & $0.16_{-0.00}^{+0.00}$ & $2.69_{-0.03}^{+0.03}$ & $0.01_{-0.00}^{+0.00}$\\
S2 & $4.79_{-0.05}^{+0.05}$ & $2.31_{-0.02}^{+0.02}$ & $14.27_{-0.02}^{+0.02}$ & $5.01_{-0.03}^{+0.04}$ & $13.82_{-0.03}^{+0.02}$ & $0.96_{-0.00}^{+0.00}$ & $15.36_{-0.14}^{+0.13}$ & $1.09_{-0.12}^{+0.11}$ & $0.72_{-0.01}^{+0.01}$ & $0.22_{-0.00}^{+0.00}$ & $3.59_{-0.05}^{+0.05}$ & $0.01_{-0.00}^{+0.00}$\\
Hb1 & $15.13_{-0.32}^{+0.34}$ & $2.25_{-0.03}^{+0.02}$ & $11.08_{-0.03}^{+0.03}$ & $3.89_{-0.05}^{+0.05}$ & $10.95_{-0.03}^{+0.03}$ & $0.99_{-0.00}^{+0.00}$ & $15.99_{-0.15}^{+0.15}$ & $4.91_{-0.18}^{+0.18}$ & $0.41_{-0.00}^{+0.00}$ & $0.12_{-0.00}^{+0.00}$ & $2.03_{-0.02}^{+0.02}$ & $0.00_{-0.00}^{+0.00}$\\
\hline
\multicolumn{13}{c}{(6) PSR J1946+3417: $M_{\rm G} = 1.832_{-0.028}^{+0.028} M_\odot$, $\nu = 315.5$~Hz}\\
\hline
N1 & $13.34_{-0.32}^{+0.32}$ & $2.11_{-0.04}^{+0.04}$ & $11.33_{-0.04}^{+0.04}$ & $4.19_{-0.08}^{+0.08}$ & $11.18_{-0.03}^{+0.03}$ & $0.98_{-0.00}^{+0.00}$ & $15.17_{-0.24}^{+0.23}$ & $3.84_{-0.28}^{+0.27}$ & $0.38_{-0.01}^{+0.01}$ & $0.13_{-0.00}^{+0.00}$ & $1.93_{-0.04}^{+0.04}$ & $0.00_{-0.00}^{+0.00}$\\
N2 & $4.82_{-0.06}^{+0.06}$ & $2.04_{-0.04}^{+0.03}$ & $15.39_{-0.01}^{+0.01}$ & $5.69_{-0.08}^{+0.08}$ & $14.78_{-0.01}^{+0.01}$ & $0.96_{-0.00}^{+0.00}$ & $15.39_{-0.01}^{+0.01}$ & $0.00_{-0.00}^{+0.00}$ & $0.66_{-0.02}^{+0.01}$ & $0.22_{-0.00}^{+0.00}$ & $3.31_{-0.08}^{+0.07}$ & $0.01_{-0.00}^{+0.00}$\\
N3 & $7.32_{-0.16}^{+0.16}$ & $2.05_{-0.04}^{+0.04}$ & $14.38_{-0.04}^{+0.03}$ & $5.31_{-0.09}^{+0.10}$ & $13.94_{-0.03}^{+0.02}$ & $0.97_{-0.00}^{+0.00}$ & $14.94_{-0.59}^{+0.00}$ & $0.56_{-0.56}^{+0.00}$ & $0.55_{-0.01}^{+0.01}$ & $0.19_{-0.00}^{+0.00}$ & $2.78_{-0.05}^{+0.05}$ & $0.01_{-0.00}^{+0.00}$\\
Hp1 & $9.35_{-0.32}^{+0.34}$ & $2.07_{-0.04}^{+0.04}$ & $13.20_{-0.05}^{+0.04}$ & $4.88_{-0.09}^{+0.09}$ & $12.90_{-0.04}^{+0.03}$ & $0.97_{-0.00}^{+0.00}$ & $15.01_{-0.23}^{+0.23}$ & $1.81_{-0.28}^{+0.27}$ & $0.49_{-0.01}^{+0.01}$ & $0.17_{-0.00}^{+0.00}$ & $2.47_{-0.04}^{+0.04}$ & $0.01_{-0.00}^{+0.00}$\\
Hp2 & $ 9.07_{-0.46}^{+0.54}$ & $2.05_{-0.04}^{+0.04}$ & $14.13_{-0.11}^{+0.09}$ & $5.22_{-0.12}^{+0.11}$ & $13.73_{-0.09}^{+0.07}$ & $0.97_{-0.00}^{+0.00}$ & $15.15_{-0.43}^{+0.13}$ & $1.02_{-0.52}^{+0.24}$ & $0.53_{-0.01}^{+0.00}$ & $0.18_{-0.00}^{+0.00}$ & $2.67_{-0.03}^{+0.02}$ & $0.01_{-0.00}^{+0.00}$\\
S1 & $8.20_{-0.23}^{+0.26}$ & $2.33_{-0.04}^{+0.04}$ & $12.15_{-0.02}^{+0.01}$ & $4.49_{-0.06}^{+0.06}$ & $11.91_{-0.02}^{+0.02}$ & $0.98_{-0.00}^{+0.00}$ & $14.87_{-0.23}^{+0.24}$ & $2.72_{-0.22}^{+0.22}$ & $0.50_{-0.01}^{+0.01}$ & $0.17_{-0.00}^{+0.00}$ & $2.53_{-0.05}^{+0.05}$ & $0.01_{-0.00}^{+0.00}$\\
S2 & $4.54_{-0.07}^{+0.06}$ & $2.18_{-0.04}^{+0.04}$ & $14.13_{-0.05}^{+0.04}$ & $5.22_{-0.06}^{+0.07}$ & $13.67_{-0.05}^{+0.04}$ & $0.96_{-0.00}^{+0.00}$ & $14.57_{-0.27}^{+0.23}$ & $0.44_{-0.22}^{+0.19}$ & $0.66_{-0.02}^{+0.01}$ & $0.22_{-0.00}^{+0.00}$ & $3.32_{-0.09}^{+0.07}$ & $0.01_{-0.00}^{+0.00}$\\
Hb1 & $13.62_{-0.35}^{+0.39}$ & $2.12_{-0.04}^{+0.04}$ & $11.23_{-0.04}^{+0.03}$ & $4.15_{-0.08}^{+0.08}$ & $11.08_{-0.03}^{+0.03}$ & $0.98_{-0.00}^{+0.00}$ & $15.17_{-0.24}^{+0.25}$ & $3.94_{-0.27}^{+0.28}$ & $0.38_{-0.01}^{+0.01}$ & $0.13_{-0.00}^{+0.00}$ & $1.92_{-0.04}^{+0.04}$ & $0.00_{-0.00}^{+0.00}$\\
\hline
\multicolumn{13}{c}{(7) PSR J1911-5958A: $M_{\rm G} = 1.33_{-0.11}^{+0.11} M_\odot$, $\nu = 305.8$~Hz}\\
\hline
N1 & $9.38_{-0.63}^{+0.68}$ & $1.46_{-0.13}^{+0.14}$ & $11.73_{-0.06}^{+0.06}$ & $5.97_{-0.48}^{+0.57}$ & $11.49_{-0.04}^{+0.03}$ & $0.98_{-0.00}^{+0.00}$ & $11.73_{-0.00}^{+0.22}$ & $0.00_{-0.00}^{+0.28}$ & $0.24_{-0.03}^{+0.03}$ & $0.16_{-0.01}^{+0.01}$ & $1.27_{-0.14}^{+0.14}$ & $0.01_{-0.00}^{+0.00}$\\
N2 & $3.89_{-0.16}^{+0.17}$ & $1.44_{-0.13}^{+0.13}$ & $15.02_{-0.10}^{+0.09}$ & $7.64_{-0.54}^{+0.64}$ & $14.30_{-0.15}^{+0.14}$ & $0.95_{-0.00}^{+0.00}$ & $15.02_{-0.10}^{+0.09}$ & $0.00_{-0.00}^{+0.00}$ & $0.39_{-0.05}^{+0.05}$ & $0.25_{-0.01}^{+0.01}$ & $2.04_{-0.25}^{+0.26}$ & $0.01_{-0.00}^{+0.00}$\\
N3 & $5.14_{-0.36}^{+0.39}$ & $1.43_{-0.13}^{+0.13}$ & $14.81_{-0.09}^{+0.08}$ & $7.54_{-0.62}^{+0.72}$ & $14.16_{-0.01}^{+0.01}$ & $0.95_{-0.00}^{+0.01}$ & $14.81_{-0.09}^{+0.08}$ & $0.00_{-0.00}^{+0.00}$ & $0.36_{-0.04}^{+0.04}$ & $0.23_{-0.01}^{+0.01}$ & $1.88_{-0.20}^{+0.19}$ & $0.01_{-0.00}^{+0.00}$\\
Hp1 & $5.97_{-0.37}^{+0.47}$ & $1.45_{-0.13}^{+0.13}$ & $13.44_{-0.02}^{+0.01}$ & $6.84_{-0.52}^{+0.61}$ & $13.00_{-0.05}^{+0.04}$ & $0.96_{-0.00}^{+0.00}$ & $13.44_{-0.02}^{+0.01}$ & $0.00_{-0.00}^{+0.00}$ & $0.31_{-0.04}^{+0.04}$ & $0.20_{-0.01}^{+0.01}$ & $1.63_{-0.19}^{+0.20}$ & $0.01_{-0.00}^{+0.00}$\\
Hp2 & $5.14_{-0.37}^{+0.39}$ & $1.43_{-0.13}^{+0.13}$ & $14.81_{-0.09}^{+0.08}$ & $7.54_{-0.62}^{+0.72}$ & $14.16_{-0.01}^{+0.02}$ & $0.95_{-0.00}^{+0.01}$ & $14.81_{-0.09}^{+0.08}$ & $0.00_{-0.00}^{+0.00}$ & $0.36_{-0.04}^{+0.04}$ & $0.23_{-0.01}^{+0.01}$ & $1.88_{-0.20}^{+0.19}$ & $0.01_{-0.00}^{+0.00}$\\
S1 & $5.77_{-0.29}^{+0.34}$ & $1.62_{-0.15}^{+0.15}$ & $11.42_{-0.24}^{+0.22}$ & $5.81_{-0.34}^{+0.39}$ & $11.15_{-0.27}^{+0.22}$ & $0.97_{-0.00}^{+0.00}$ & $11.42_{-0.24}^{+0.22}$ & $0.00_{-0.00}^{+0.00}$ & $0.30_{-0.04}^{+0.04}$ & $0.19_{-0.01}^{+0.01}$ & $1.56_{-0.20}^{+0.21}$ & $0.01_{-0.00}^{+0.00}$\\
S2 & $3.68_{-0.13}^{+0.15}$ & $1.53_{-0.14}^{+0.14}$ & $13.06_{-0.30}^{+0.27}$ & $6.65_{-0.38}^{+0.43}$ & $12.54_{-0.31}^{+0.34}$ & $0.96_{-0.00}^{+0.01}$ & $13.06_{-0.30}^{+0.27}$ & $0.00_{-0.00}^{+0.00}$ & $0.38_{-0.05}^{+0.05}$ & $0.25_{-0.01}^{+0.01}$ & $1.99_{-0.26}^{+0.27}$ & $0.01_{-0.00}^{+0.00}$\\
Hb1 & $9.45_{-0.64}^{+0.70}$ & $1.46_{-0.13}^{+0.14}$ & $11.57_{-0.04}^{+0.03}$ & $5.89_{-0.47}^{+0.55}$ & $11.33_{-0.02}^{+0.01}$ & $0.98_{-0.00}^{+0.00}$ & $11.57_{-0.00}^{+0.37}$ & $0.00_{-0.00}^{+0.41}$ & $0.24_{-0.03}^{+0.03}$ & $0.15_{-0.01}^{+0.01}$ & $1.25_{-0.14}^{+0.15}$ & $0.01_{-0.00}^{+0.00}$\\
\hline
\multicolumn{13}{c}{(8) PSR J0751+1807: $M_{\rm G} = 1.64_{-0.15}^{+0.15} M_\odot$, $\nu = 287.4$~Hz}\\
\hline
N1 & $11.51_{-1.11}^{+1.39}$ & $1.85_{-0.19}^{+0.20}$ & $11.51_{-0.15}^{+0.11}$ & $4.75_{-0.45}^{+0.53}$ & $11.36_{-0.13}^{+0.07}$ & $0.98_{-0.00}^{+0.00}$ & $13.62_{-1.21}^{+1.28}$ & $2.11_{-1.33}^{+1.43}$ & $0.30_{-0.04}^{+0.04}$ & $0.13_{-0.01}^{+0.01}$ & $1.67_{-0.20}^{+0.20}$ & $0.00_{-0.00}^{+0.00}$\\
N2 & $4.43_{-0.27}^{+0.31}$ & $1.80_{-0.18}^{+0.18}$ & $15.23_{-0.11}^{+0.09}$ & $6.28_{-0.49}^{+0.58}$ & $14.70_{-0.15}^{+0.11}$ & $0.96_{-0.00}^{+0.00}$ & $15.23_{-0.11}^{+0.09}$ & $0.00_{-0.00}^{+0.00}$ & $0.50_{-0.07}^{+0.07}$ & $0.21_{-0.01}^{+0.01}$ & $2.78_{-0.37}^{+0.39}$ & $0.01_{-0.00}^{+0.00}$\\
N3 & $6.37_{-0.63}^{+0.74}$ & $1.81_{-0.18}^{+0.19}$ & $14.53_{-0.15}^{+0.12}$ & $6.00_{-0.56}^{+0.66}$ & $14.11_{-0.10}^{+0.07}$ & $0.97_{-0.00}^{+0.00}$ & $14.53_{-0.00}^{+0.20}$ & $0.00_{-0.00}^{+0.35}$ & $0.44_{-0.05}^{+0.05}$ & $0.18_{-0.01}^{+0.01}$ & $2.42_{-0.27}^{+0.27}$ & $0.01_{-0.00}^{+0.00}$\\
Hp1 & $7.64_{-0.93}^{+1.28}$ & $1.83_{-0.19}^{+0.19}$ & $13.37_{-0.14}^{+0.05}$ & $5.52_{-0.51}^{+0.58}$ & $13.06_{-0.09}^{+0.01}$ & $0.97_{-0.00}^{+0.00}$ & $13.54_{-0.12}^{+1.20}$ & $0.18_{-0.18}^{+1.34}$ & $0.39_{-0.05}^{+0.04}$ & $0.17_{-0.01}^{+0.01}$ & $2.17_{-0.26}^{+0.23}$ & $0.01_{-0.00}^{+0.00}$\\
Hp2 & $6.82_{-1.03}^{+1.66}$ & $1.81_{-0.18}^{+0.19}$ & $14.50_{-0.29}^{+0.14}$ & $5.99_{-0.61}^{+0.67}$ & $14.09_{-0.23}^{+0.09}$ & $0.97_{-0.00}^{+0.00}$ & $14.50_{-0.00}^{-0.17}$ & $0.00_{-0.00}^{+0.46}$ & $0.43_{-0.05}^{+0.04}$ & $0.18_{-0.02}^{+0.01}$ & $2.41_{-0.26}^{+0.21}$ & $0.01_{-0.00}^{+0.00}$\\
S1 & $6.96_{-0.65}^{+0.93}$ & $2.05_{-0.21}^{+0.21}$ & $11.94_{-0.23}^{+0.16}$ & $4.93_{-0.35}^{+0.39}$ & $11.72_{-0.24}^{+0.17}$ & $0.98_{-0.00}^{+0.00}$ & $13.35_{-1.28}^{+1.25}$ & $1.41_{-1.06}^{+1.09}$ & $0.39_{-0.05}^{+0.05}$ & $0.16_{-0.01}^{+0.01}$ & $2.16_{-0.29}^{+0.29}$ & $0.01_{-0.00}^{+0.00}$\\
S2 & $4.16_{-0.25}^{+0.31}$ & $1.93_{-0.19}^{+0.20}$ & $13.76_{-0.32}^{+0.27}$ & $5.68_{-0.37}^{+0.43}$ & $13.34_{-0.33}^{+0.28}$ & $0.97_{-0.00}^{+0.00}$ & $13.76_{-0.32}^{+0.66}$ & $0.00_{-0.00}^{+0.38}$ & $0.50_{-0.07}^{+0.07}$ & $0.21_{-0.01}^{+0.01}$ & $2.78_{-0.40}^{+0.41}$ & $0.01_{-0.00}^{+0.00}$\\
Hb1 & $11.65_{-1.15}^{+1.49}$ & $1.86_{-0.19}^{+0.20}$ & $11.40_{-0.14}^{+0.09}$ & $4.71_{-0.45}^{+0.52}$ & $11.24_{-0.11}^{+0.06}$ & $0.98_{-0.00}^{+0.00}$ & $13.64_{-1.24}^{+1.27}$ & $2.24_{-1.33}^{+1.41}$ & $0.30_{-0.04}^{+0.04}$ & $0.13_{-0.01}^{+0.01}$ & $1.66_{-0.20}^{+0.20}$ & $0.00_{-0.00}^{+0.00}$\\
\hline
\multicolumn{13}{c}{(9) PSR J2234+0611: $M_{\rm G} = 1.393_{-0.013}^{+0.013} M_\odot$, $\nu = 279.3$~Hz}\\
\hline
N1 & $9.79_{-0.08}^{+0.08}$ & $1.54_{-0.02}^{+0.02}$ & $11.68_{-0.01}^{+0.01}$ & $5.67_{-0.06}^{+0.06}$ & $11.48_{-0.01}^{+0.01}$ & $0.98_{-0.00}^{+0.00}$ & $11.68_{-0.00}^{+0.06}$ & $0.00_{-0.00}^{+0.06}$ & $0.24_{-0.00}^{+0.00}$ & $0.14_{-0.00}^{+0.00}$ & $1.35_{-0.02}^{+0.02}$ & $0.01_{-0.00}^{+0.00}$\\
N2 & $4.00_{-0.02}^{+0.02}$ & $1.51_{-0.02}^{+0.02}$ & $15.01_{-0.01}^{+0.01}$ & $7.30_{-0.06}^{+0.06}$ & $14.44_{-0.02}^{+0.02}$ & $0.96_{-0.00}^{+0.00}$ & $15.01_{-0.01}^{+0.01}$ & $0.00_{-0.00}^{+0.00}$ & $0.38_{-0.01}^{+0.01}$ & $0.22_{-0.00}^{+0.00}$ & $2.18_{-0.03}^{+0.03}$ & $0.01_{-0.00}^{+0.00}$\\
N3 & $5.38_{-0.05}^{+0.05}$ & $1.51_{-0.02}^{+0.02}$ & $14.70_{-0.01}^{+0.01}$ & $7.15_{-0.07}^{+0.07}$ & $14.21_{-0.00}^{+0.00}$ & $0.96_{-0.00}^{+0.00}$ & $14.70_{-0.01}^{+0.01}$ & $0.00_{-0.00}^{+0.00}$ & $0.35_{-0.00}^{+0.00}$ & $0.20_{-0.00}^{+0.00}$ & $1.97_{-0.02}^{+0.02}$ & $0.01_{-0.00}^{+0.00}$\\
Hp1 & $6.24_{-0.06}^{+0.06}$ & $1.52_{-0.02}^{+0.02}$ & $13.41_{-0.00}^{+0.00}$ & $6.52_{-0.06}^{+0.06}$ & $13.05_{-0.00}^{+0.00}$ & $0.97_{-0.00}^{+0.00}$ & $13.41_{-0.00}^{+0.00}$ & $0.00_{-0.00}^{+0.00}$ & $0.30_{-0.00}^{+0.00}$ & $0.18_{-0.00}^{+0.00}$ & $1.73_{-0.02}^{+0.02}$ & $0.01_{-0.00}^{+0.00}$\\
Hp2 & $5.38_{-0.04}^{+0.05}$ & $1.51_{-0.01}^{+0.02}$ & $14.70_{-0.01}^{+0.01}$ & $7.15_{-0.07}^{+0.07}$ & $14.21_{-0.00}^{+0.00}$ & $0.96_{-0.00}^{+0.00}$ & $14.70_{-0.01}^{+0.01}$ & $0.00_{-0.00}^{+0.00}$ & $0.35_{-0.00}^{+0.00}$ & $0.20_{-0.00}^{+0.00}$ & $1.97_{-0.02}^{+0.02}$ & $0.01_{-0.00}^{+0.00}$\\
S1 & $5.98_{-0.04}^{+0.04}$ & $1.71_{-0.02}^{+0.02}$ & $11.53_{-0.03}^{+0.02}$ & $5.60_{-0.04}^{+0.04}$ & $11.31_{-0.03}^{+0.03}$ & $0.98_{-0.00}^{+0.00}$ & $11.53_{-0.03}^{+0.02}$ & $0.00_{-0.00}^{+0.00}$ & $0.29_{-0.00}^{+0.00}$ & $0.17_{-0.00}^{+0.00}$ & $1.68_{-0.02}^{+0.03}$ & $0.01_{-0.00}^{+0.00}$\\
S2 & $3.77_{-0.02}^{+0.02}$ & $1.61_{-0.02}^{+0.02}$ & $13.19_{-0.03}^{+0.03}$ & $6.41_{-0.04}^{+0.04}$ & $12.78_{-0.03}^{+0.03}$ & $0.97_{-0.00}^{+0.00}$ & $13.19_{-0.03}^{+0.03}$ & $0.00_{-0.00}^{+0.00}$ & $0.38_{-0.01}^{+0.01}$ & $0.22_{-0.00}^{+0.00}$ & $2.14_{-0.03}^{+0.03}$ & $0.01_{-0.00}^{+0.00}$\\
Hb1 & $9.86_{-0.08}^{+0.08}$ & $1.54_{-0.01}^{+0.02}$ & $11.53_{-0.00}^{+0.00}$ & $5.60_{-0.05}^{+0.06}$ & $11.34_{-0.00}^{+0.00}$ & $0.98_{-0.00}^{+0.00}$ & $11.62_{-0.08}^{+0.11}$ & $0.09_{-0.09}^{+0.11}$ & $0.23_{-0.00}^{+0.00}$ & $0.14_{-0.00}^{+0.00}$ & $1.33_{-0.02}^{+0.02}$ & $0.00_{-0.00}^{+0.00}$\\
\hline
\multicolumn{13}{c}{(10) PSR J1807-2500B: $M_{\rm G} = 1.3655_{-0.0021}^{+0.0021} M_\odot$, $\nu = 238.7$~Hz}\\
\hline
N1 & $9.64_{-0.01}^{+0.01}$ & $1.51_{-0.00}^{+0.00}$ & $11.66_{-0.00}^{+0.00}$ & $5.78_{-0.01}^{+0.01}$ & $11.51_{-0.00}^{+0.00}$ & $0.99_{-0.00}^{+0.00}$ & $11.66_{-0.00}^{+0.00}$ & $0.00_{-0.00}^{+0.00}$ & $0.20_{-0.00}^{+0.00}$ & $0.12_{-0.00}^{+0.00}$ & $1.31_{-0.00}^{+0.00}$ & $0.00_{-0.00}^{+0.00}$\\
N2 & $3.97_{-0.00}^{+0.00}$ & $1.48_{-0.00}^{+0.00}$ & $14.90_{-0.00}^{+0.00}$ & $7.39_{-0.01}^{+0.01}$ & $14.50_{-0.00}^{+0.00}$ & $0.97_{-0.00}^{+0.00}$ & $14.90_{-0.00}^{+0.00}$ & $0.00_{-0.00}^{+0.00}$ & $0.31_{-0.00}^{+0.00}$ & $0.19_{-0.00}^{+0.00}$ & $2.09_{-0.01}^{+0.00}$ & $0.01_{-0.00}^{+0.00}$\\
N3 & $5.31_{-0.01}^{+0.01}$ & $1.48_{-0.00}^{+0.00}$ & $14.64_{-0.00}^{+0.00}$ & $7.26_{-0.01}^{+0.01}$ & $14.28_{-0.00}^{+0.00}$ & $0.97_{-0.00}^{+0.00}$ & $14.64_{-0.00}^{+0.00}$ & $0.00_{-0.00}^{+0.00}$ & $0.29_{-0.00}^{+0.00}$ & $0.17_{-0.00}^{+0.00}$ & $1.91_{-0.00}^{+0.00}$ & $0.01_{-0.00}^{+0.00}$\\
Hp1 & $6.15_{-0.01}^{+0.01}$ & $1.49_{-0.00}^{+0.00}$ & $13.36_{-0.00}^{+0.00}$ & $6.62_{-0.01}^{+0.01}$ & $13.09_{-0.00}^{+0.00}$ & $0.98_{-0.00}^{+0.00}$ & $13.36_{-0.00}^{+0.00}$ & $0.00_{-0.00}^{+0.00}$ & $0.25_{-0.00}^{+0.00}$ & $0.15_{-0.00}^{+0.00}$ & $1.68_{-0.00}^{+0.00}$ & $0.01_{-0.00}^{+0.00}$\\
Hp2 & $5.31_{-0.01}^{+0.01}$ & $1.48_{-0.00}^{+0.00}$ & $14.63_{-0.00}^{+0.00}$ & $7.26_{-0.01}^{+0.01}$ & $14.28_{-0.00}^{+0.00}$ & $0.97_{-0.00}^{+0.00}$ & $14.63_{-0.00}^{+0.00}$ & $0.00_{-0.00}^{+0.00}$ & $0.29_{-0.00}^{+0.00}$ & $0.17_{-0.00}^{+0.00}$ & $1.91_{-0.00}^{+0.00}$ & $0.01_{-0.00}^{+0.00}$\\
S1 & $5.91_{-0.01}^{+0.01}$ & $1.68_{-0.00}^{+0.00}$ & $11.44_{-0.00}^{+0.00}$ & $5.67_{-0.01}^{+0.01}$ & $11.35_{-0.00}^{+0.00}$ & $0.99_{-0.00}^{+0.00}$ & $11.44_{-0.00}^{+0.00}$ & $0.00_{-0.00}^{+0.00}$ & $0.24_{-0.00}^{+0.00}$ & $0.15_{-0.00}^{+0.00}$ & $1.61_{-0.00}^{+0.00}$ & $0.01_{-0.00}^{+0.00}$\\
S2 & $3.75_{-0.00}^{+0.00}$ & $1.58_{-0.00}^{+0.00}$ & $13.08_{-0.01}^{+0.01}$ & $6.48_{-0.01}^{+0.01}$ & $12.79_{-0.01}^{+0.01}$ & $0.98_{-0.00}^{+0.00}$ & $13.08_{-0.01}^{+0.01}$ & $0.00_{-0.00}^{+0.00}$ & $0.31_{-0.00}^{+0.00}$ & $0.19_{-0.00}^{+0.00}$ & $2.06_{-0.01}^{+0.01}$ & $0.01_{-0.00}^{+0.00}$\\
Hb1 & $9.71_{-0.01}^{+0.01}$ & $1.51_{-0.00}^{+0.00}$ & $11.51_{-0.00}^{+0.00}$ & $5.71_{-0.01}^{+0.01}$ & $11.37_{-0.00}^{+0.00}$ & $0.99_{-0.00}^{+0.00}$ & $11.51_{-0.00}^{+0.00}$ & $0.00_{-0.00}^{+0.00}$ & $0.19_{-0.00}^{+0.00}$ & $0.12_{-0.00}^{+0.00}$ & $1.29_{-0.00}^{+0.00}$ & $0.00_{-0.00}^{+0.00}$\\
\hline
\end{tabular}}
\begin{flushleft}
Table 3 is continued on the next page.
\end{flushleft}
\end{table*}

\begin{table*}
\centering
{\bf Table~\ref{table_result}} (continued)\\
\scalebox{0.80}{%
\begin{tabular}{l@{}c@{\hspace{2mm}}c@{\hspace{2mm}}c@{\hspace{2mm}}c@{\hspace{2mm}}c@{\hspace{2mm}}c@{\hspace{2mm}}c@{\hspace{2mm}}c@{\hspace{2mm}}c@{\hspace{2mm}}c@{\hspace{2mm}}c@{\hspace{2mm}}c}
\hline
  EoS\footnotemark[1] & $\rho_{\rm c}$\footnotemark[2] & $M_{\rm 0}$\footnotemark[3] & $R_{\rm e}$\footnotemark[4] & $R_{\rm e}/r_{\rm g}$\footnotemark[5] & $R_{\rm p}$\footnotemark[6] & $R_{\rm p}/R_{\rm e}$ & $r_{\rm orb}$\footnotemark[7] & $r_{\rm orb}-R_{\rm e}$ & $J$\footnotemark[8] & $cJ/GM_{\rm G}^2$\footnotemark[9] & $I$\footnotemark[10] & $T/W$\footnotemark[11]\\
 & ($10^{14}$ & ($M_{\odot}$) & (km) & & (km) & & (km) & (km) & ($10^{49}$ & & ($10^{45}$ & \\
 & g cm$^{-3}$) & & & & & & & & g cm$^2$ s$^{-1}$) & & g cm$^2$) & \\
\hline
\multicolumn{13}{c}{(11) PSR J1713+0747: $M_{\rm G} = 1.33_{-0.08}^{+0.09} M_\odot$, $\nu = 218.9$~Hz}\\
\hline
N1 & $9.43_{-0.47}^{+0.55}$ & $1.46_{-0.10}^{+0.11}$ & $11.67_{-0.04}^{+0.03}$ & $5.94_{-0.40}^{+0.40}$ & $11.54_{-0.04}^{+0.03}$ & $0.99_{-0.00}^{+0.00}$ & $11.67_{-0.00}^{+0.30}$ & $0.00_{-0.00}^{+0.35}$ & $0.17_{-0.01}^{+0.02}$ & $0.11_{-0.00}^{+0.00}$ & $1.26_{-0.10}^{+0.12}$ & $0.00_{-0.00}^{+0.00}$\\
N2 & $3.92_{-0.12}^{+0.14}$ & $1.44_{-0.09}^{+0.10}$ & $14.83_{-0.09}^{+0.10}$ & $7.55_{-0.43}^{+0.44}$ & $14.49_{-0.11}^{+0.09}$ & $0.97_{-0.00}^{+0.00}$ & $14.83_{-0.09}^{+0.10}$ & $0.00_{-0.00}^{+0.00}$ & $0.28_{-0.02}^{+0.03}$ & $0.18_{-0.01}^{+0.01}$ & $2.01_{-0.18}^{+0.21}$ & $0.01_{-0.00}^{+0.00}$\\
N3 & $5.19_{-0.27}^{+0.33}$ & $1.44_{-0.09}^{+0.11}$ & $14.62_{-0.05}^{+0.05}$ & $7.45_{-0.50}^{+0.50}$ & $14.31_{-0.02}^{+0.00}$ & $0.98_{-0.00}^{+0.00}$ & $14.62_{-0.05}^{+0.05}$ & $0.00_{-0.00}^{+0.00}$ & $0.25_{-0.02}^{+0.02}$ & $0.16_{-0.01}^{+0.01}$ & $1.84_{-0.14}^{+0.16}$ & $0.01_{-0.00}^{+0.00}$\\
Hp1 & $6.01_{-0.29}^{+0.39}$ & $1.45_{-0.09}^{+0.11}$ & $13.33_{-0.02}^{+0.01}$ & $6.78_{-0.42}^{+0.42}$ & $13.10_{-0.03}^{+0.03}$ & $0.98_{-0.00}^{+0.00}$ & $13.33_{-0.02}^{+0.01}$ & $0.00_{-0.00}^{+0.00}$ & $0.22_{-0.02}^{+0.02}$ & $0.14_{-0.01}^{+0.00}$ & $1.61_{-0.14}^{+0.16}$ & $0.00_{-0.00}^{+0.00}$\\
Hp2 & $5.20_{-0.27}^{+0.32}$ & $1.44_{-0.09}^{+0.11}$ & $14.62_{-0.05}^{+0.05}$ & $7.44_{-0.49}^{+0.50}$ & $14.31_{-0.02}^{+0.00}$ & $0.98_{-0.00}^{+0.00}$ & $14.62_{-0.05}^{+0.05}$ & $0.00_{-0.00}^{+0.00}$ & $0.25_{-0.02}^{+0.02}$ & $0.16_{-0.01}^{+0.01}$ & $1.84_{-0.14}^{+0.16}$ & $0.01_{-0.00}^{+0.00}$\\
S1 & $5.81_{-0.22}^{+0.28}$ & $1.63_{-0.11}^{+0.12}$ & $11.36_{-0.18}^{+0.18}$ & $5.78_{-0.28}^{+0.27}$ & $11.28_{-0.18}^{+0.18}$ & $0.99_{-0.00}^{+0.00}$ & $11.36_{-0.18}^{+0.31}$ & $0.00_{-0.00}^{+0.13}$ & $0.21_{-0.02}^{+0.02}$ & $0.14_{-0.00}^{+0.00}$ & $1.54_{-0.15}^{+0.17}$ & $0.00_{-0.00}^{+0.00}$\\
S2 & $3.70_{-0.10}^{+0.12}$ & $1.53_{-0.10}^{+0.11}$ & $12.94_{-0.22}^{+0.23}$ & $6.59_{-0.31}^{+0.30}$ & $12.80_{-0.22}^{+0.23}$ & $0.99_{-0.00}^{+0.00}$ & $12.94_{-0.22}^{+0.23}$ & $0.00_{-0.00}^{+0.00}$ & $0.27_{-0.03}^{+0.03}$ & $0.17_{-0.00}^{+0.00}$ & $1.95_{-0.19}^{+0.22}$ & $0.01_{-0.00}^{+0.00}$\\
Hb1 & $9.50_{-0.47}^{+0.56}$ & $1.46_{-0.10}^{+0.11}$ & $11.51_{-0.03}^{+0.02}$ & $5.86_{-0.38}^{+0.39}$ & $11.38_{-0.02}^{+0.01}$ & $0.99_{-0.00}^{+0.00}$ & $11.51_{-0.00}^{+0.46}$ & $0.00_{-0.00}^{+0.49}$ & $0.17_{-0.01}^{+0.02}$ & $0.11_{-0.00}^{+0.00}$ & $1.24_{-0.10}^{+0.12}$ & $0.00_{-0.00}^{+0.00}$\\
\hline
\multicolumn{13}{c}{(12) PSR J1012+5307: $M_{\rm G} = 1.83_{-0.11}^{+0.11} M_\odot$, $\nu = 190.1$~Hz}\\
\hline
N1 & $13.41_{-1.13}^{+1.46}$ & $2.11_{-0.15}^{+0.16}$ & $11.28_{-0.16}^{+0.12}$ & $4.17_{-0.29}^{+0.31}$ & $11.22_{-0.15}^{+0.11}$ & $0.99_{-0.00}^{+0.00}$ & $15.57_{-0.95}^{+0.96}$ & $4.29_{-1.07}^{+1.11}$ & $0.23_{-0.02}^{+0.02}$ & $0.08_{-0.00}^{+0.00}$ & $1.92_{-0.15}^{+0.14}$ & $0.00_{-0.00}^{+0.00}$\\
N2 & $4.88_{-0.24}^{+0.28}$ & $2.04_{-0.14}^{+0.14}$ & $15.21_{-0.07}^{+0.03}$ & $5.63_{-0.31}^{+0.34}$ & $14.95_{-0.04}^{+0.05}$ & $0.98_{-0.00}^{+0.00}$ & $15.46_{-0.31}^{+0.76}$ & $0.25_{-0.25}^{+0.73}$ & $0.39_{-0.04}^{+0.03}$ & $0.13_{-0.00}^{+0.00}$ & $3.26_{-0.30}^{+0.28}$ & $0.00_{-0.00}^{+0.00}$\\
N3 & $7.40_{-0.59}^{+0.75}$ & $2.05_{-0.14}^{+0.14}$ & $14.22_{-0.15}^{+0.11}$ & $5.26_{-0.35}^{+0.38}$ & $14.06_{-0.11}^{+0.10}$ & $0.99_{-0.00}^{+0.00}$ & $15.42_{-0.94}^{+0.87}$ & $1.20_{-1.05}^{+1.03}$ & $0.33_{-0.02}^{+0.02}$ & $0.11_{-0.01}^{+0.01}$ & $2.74_{-0.20}^{+0.18}$ & $0.00_{-0.00}^{+0.00}$\\
Hp1 & $9.49_{-1.13}^{+1.66}$ & $2.07_{-0.14}^{+0.15}$ & $13.09_{-0.22}^{+0.13}$ & $4.84_{-0.35}^{+0.36}$ & $12.98_{-0.21}^{+0.12}$ & $0.99_{-0.00}^{+0.00}$ & $15.44_{-0.95}^{+0.96}$ & $2.35_{-1.08}^{+1.18}$ & $0.29_{-0.02}^{+0.02}$ & $0.10_{-0.01}^{+0.01}$ & $2.43_{-0.16}^{+0.13}$ & $0.00_{-0.00}^{+0.00}$\\
Hp2 & $9.29_{-1.56}^{+3.34}$ & $2.05_{-0.14}^{+0.14}$ & $13.96_{-0.60}^{+0.28}$ & $5.16_{-0.50}^{+0.44}$ & $13.81_{-0.55}^{+0.26}$ & $0.99_{-0.00}^{+0.00}$ & $15.42_{-0.93}^{+0.95}$ & $1.46_{-1.21}^{+1.55}$ & $0.31_{-0.02}^{+0.00}$ & $0.11_{-0.01}^{+0.01}$ & $2.62_{-0.13}^{+0.00}$ & $0.00_{-0.00}^{+0.00}$\\
S1 & $8.34_{-0.85}^{+1.29}$ & $2.33_{-0.16}^{+0.16}$ & $12.07_{-0.09}^{+0.02}$ & $4.46_{-0.24}^{+0.25}$ & $12.02_{-0.09}^{+0.02}$ & $0.99_{-0.00}^{+0.00}$ & $15.32_{-0.96}^{+0.96}$ & $3.25_{-0.87}^{+0.94}$ & $0.30_{-0.02}^{+0.02}$ & $0.10_{-0.00}^{+0.00}$ & $2.49_{-0.20}^{+0.18}$ & $0.00_{-0.00}^{+0.00}$\\
S2 & $4.61_{-0.25}^{+0.30}$ & $2.18_{-0.15}^{+0.15}$ & $13.99_{-0.19}^{+0.16}$ & $5.17_{-0.24}^{+0.26}$ & $13.89_{-0.19}^{+0.16}$ & $0.99_{-0.00}^{+0.00}$ & $14.99_{-0.93}^{+0.93}$ & $0.99_{-0.75}^{+0.77}$ & $0.39_{-0.04}^{+0.04}$ & $0.13_{-0.00}^{+0.00}$ & $3.25_{-0.30}^{+0.30}$ & $0.00_{-0.00}^{+0.00}$\\
Hb1 & $13.70_{-1.25}^{+1.80}$ & $2.12_{-0.15}^{+0.16}$ & $11.18_{-0.17}^{+0.11}$ & $4.13_{-0.29}^{+0.31}$ & $11.12_{-0.16}^{+0.11}$ & $0.99_{-0.00}^{+0.00}$ & $15.57_{-0.95}^{+0.96}$ & $4.39_{-1.06}^{+1.13}$ & $0.23_{-0.02}^{+0.02}$ & $0.08_{-0.00}^{+0.00}$ & $1.90_{-0.14}^{+0.14}$ & $0.00_{-0.00}^{+0.00}$\\
\hline
\multicolumn{13}{c}{(13) PSR B1855+09: $M_{\rm G} = 1.30_{-0.10}^{+0.11} M_\odot$, $\nu = 186.6$~Hz}\\
\hline
N1 & $9.26_{-0.57}^{+0.67}$ & $1.42_{-0.12}^{+0.14}$ & $11.66_{-0.05}^{+0.04}$ & $6.07_{-0.50}^{+0.53}$ & $11.56_{-0.04}^{+0.03}$ & $0.99_{-0.00}^{+0.00}$ & $11.66_{-0.00}^{+0.31}$ & $0.00_{-0.00}^{+0.36}$ & $0.14_{-0.01}^{+0.02}$ & $0.10_{-0.00}^{+0.01}$ & $1.23_{-0.13}^{+0.14}$ & $0.00_{-0.00}^{+0.00}$\\
N2 & $3.89_{-0.15}^{+0.17}$ & $1.40_{-0.12}^{+0.13}$ & $14.76_{-0.12}^{+0.12}$ & $7.68_{-0.54}^{+0.58}$ & $14.51_{-0.14}^{+0.11}$ & $0.98_{-0.00}^{+0.00}$ & $14.76_{-0.12}^{+0.12}$ & $0.00_{-0.00}^{+0.00}$ & $0.23_{-0.03}^{+0.03}$ & $0.15_{-0.01}^{+0.01}$ & $1.93_{-0.22}^{+0.26}$ & $0.01_{-0.00}^{+0.00}$\\
N3 & $5.11_{-0.33}^{+0.39}$ & $1.40_{-0.12}^{+0.13}$ & $14.58_{-0.05}^{+0.05}$ & $7.59_{-0.61}^{+0.67}$ & $14.35_{-0.03}^{+0.01}$ & $0.98_{-0.00}^{+0.00}$ & $14.58_{-0.05}^{+0.05}$ & $0.00_{-0.00}^{+0.00}$ & $0.21_{-0.02}^{+0.02}$ & $0.14_{-0.01}^{+0.01}$ & $1.78_{-0.17}^{+0.19}$ & $0.00_{-0.00}^{+0.00}$\\
Hp1 & $5.91_{-0.33}^{+0.46}$ & $1.41_{-0.12}^{+0.13}$ & $13.29_{-0.03}^{+0.02}$ & $6.92_{-0.53}^{+0.56}$ & $13.12_{-0.04}^{+0.03}$ & $0.99_{-0.00}^{+0.00}$ & $13.29_{-0.03}^{+0.02}$ & $0.00_{-0.00}^{+0.00}$ & $0.18_{-0.02}^{+0.02}$ & $0.12_{-0.01}^{+0.01}$ & $1.55_{-0.17}^{+0.19}$ & $0.00_{-0.00}^{+0.00}$\\
Hp2 & $5.10_{-0.33}^{+0.39}$ & $1.40_{-0.12}^{+0.13}$ & $14.58_{-0.05}^{+0.05}$ & $7.60_{-0.62}^{+0.66}$ & $14.35_{-0.03}^{+0.01}$ & $0.98_{-0.00}^{+0.00}$ & $14.58_{-0.05}^{+0.05}$ & $0.00_{-0.00}^{+0.00}$ & $0.21_{-0.02}^{+0.02}$ & $0.14_{-0.01}^{+0.01}$ & $1.78_{-0.17}^{+0.19}$ & $0.00_{-0.00}^{+0.00}$\\
S1 & $5.74_{-0.26}^{+0.33}$ & $1.59_{-0.13}^{+0.15}$ & $11.29_{-0.23}^{+0.22}$ & $5.88_{-0.35}^{+0.36}$ & $11.24_{-0.23}^{+0.23}$ & $0.99_{-0.00}^{+0.00}$ & $11.29_{-0.23}^{+0.42}$ & $0.00_{-0.00}^{+0.20}$ & $0.17_{-0.02}^{+0.02}$ & $0.12_{-0.00}^{+0.00}$ & $1.48_{-0.18}^{+0.20}$ & $0.00_{-0.00}^{+0.00}$\\
S2 & $3.67_{-0.12}^{+0.15}$ & $1.49_{-0.12}^{+0.14}$ & $12.85_{-0.28}^{+0.29}$ & $6.69_{-0.39}^{+0.40}$ & $12.75_{-0.28}^{+0.29}$ & $0.99_{-0.00}^{+0.00}$ & $12.85_{-0.28}^{+0.29}$ & $0.00_{-0.00}^{+0.00}$ & $0.22_{-0.03}^{+0.03}$ & $0.15_{-0.00}^{+0.00}$ & $1.87_{-0.23}^{+0.27}$ & $0.01_{-0.00}^{+0.00}$\\
Hb1 & $9.33_{-0.57}^{+0.68}$ & $1.43_{-0.12}^{+0.14}$ & $11.50_{-0.03}^{+0.02}$ & $5.99_{-0.48}^{+0.51}$ & $11.40_{-0.02}^{+0.01}$ & $0.99_{-0.00}^{+0.00}$ & $11.50_{-0.00}^{+0.46}$ & $0.00_{-0.00}^{+0.49}$ & $0.14_{-0.02}^{+0.02}$ & $0.09_{-0.00}^{+0.00}$ & $1.20_{-0.13}^{+0.14}$ & $0.00_{-0.00}^{+0.00}$\\
\hline
\multicolumn{13}{c}{(14) PSR J0437-4715: $M_{\rm G} = 1.44_{-0.07}^{+0.07} M_\odot$, $\nu = 173.6$~Hz}\\
\hline
N1 & $10.12_{-0.43}^{+0.47}$ & $1.60_{-0.09}^{+0.09}$ & $11.59_{-0.04}^{+0.03}$ & $5.45_{-0.27}^{+0.29}$ & $11.51_{-0.03}^{+0.03}$ & $0.99_{-0.00}^{+0.00}$ & $12.26_{-0.60}^{+0.59}$ & $0.67_{-0.63}^{+0.63}$ & $0.15_{-0.01}^{+0.01}$ & $0.08_{-0.00}^{+0.00}$ & $1.40_{-0.09}^{+0.09}$ & $0.00_{-0.00}^{+0.00}$\\
N2 & $4.11_{-0.11}^{+0.12}$ & $1.57_{-0.08}^{+0.08}$ & $14.89_{-0.07}^{+0.06}$ & $7.00_{-0.30}^{+0.33}$ & $14.67_{-0.08}^{+0.08}$ & $0.98_{-0.00}^{+0.00}$ & $14.89_{-0.07}^{+0.06}$ & $0.00_{-0.00}^{+0.00}$ & $0.25_{-0.02}^{+0.02}$ & $0.13_{-0.00}^{+0.00}$ & $2.25_{-0.17}^{+0.17}$ & $0.00_{-0.00}^{+0.00}$\\
N3 & $5.62_{-0.26}^{+0.27}$ & $1.57_{-0.08}^{+0.08}$ & $14.51_{-0.04}^{+0.03}$ & $6.82_{-0.33}^{+0.37}$ & $14.33_{-0.03}^{+0.03}$ & $0.99_{-0.00}^{+0.00}$ & $14.51_{-0.04}^{+0.03}$ & $0.00_{-0.00}^{+0.00}$ & $0.22_{-0.01}^{+0.01}$ & $0.12_{-0.00}^{+0.00}$ & $2.02_{-0.13}^{+0.13}$ & $0.00_{-0.00}^{+0.00}$\\
Hp1 & $6.52_{-0.32}^{+0.37}$ & $1.58_{-0.09}^{+0.09}$ & $13.30_{-0.01}^{+0.00}$ & $6.25_{-0.29}^{+0.32}$ & $13.17_{-0.01}^{+0.01}$ & $0.99_{-0.00}^{+0.00}$ & $13.30_{-0.01}^{+0.00}$ & $0.00_{-0.00}^{+0.00}$ & $0.20_{-0.01}^{+0.01}$ & $0.11_{-0.00}^{+0.00}$ & $1.80_{-0.13}^{+0.12}$ & $0.00_{-0.00}^{+0.00}$\\
Hp2 & $5.63_{-0.27}^{+0.37}$ & $1.57_{-0.08}^{+0.08}$ & $14.51_{-0.04}^{+0.03}$ & $6.82_{-0.33}^{+0.36}$ & $14.33_{-0.03}^{+0.03}$ & $0.99_{-0.00}^{+0.00}$ & $14.51_{-0.04}^{+0.03}$ & $0.00_{-0.00}^{+0.00}$ & $0.22_{-0.01}^{+0.01}$ & $0.12_{-0.00}^{+0.00}$ & $2.02_{-0.13}^{+0.12}$ & $0.00_{-0.00}^{+0.00}$\\
S1 & $6.18_{-0.23}^{+0.26}$ & $1.78_{-0.10}^{+0.10}$ & $11.57_{-0.13}^{+0.12}$ & $5.44_{-0.20}^{+0.21}$ & $11.52_{-0.14}^{+0.12}$ & $0.99_{-0.00}^{+0.00}$ & $12.03_{-0.59}^{+0.60}$ & $0.45_{-0.45}^{+0.48}$ & $0.19_{-0.01}^{+0.01}$ & $0.10_{-0.00}^{+0.00}$ & $1.74_{-0.13}^{+0.13}$ & $0.00_{-0.00}^{+0.00}$\\
S2 & $3.87_{-0.10}^{+0.11}$ & $1.67_{-0.09}^{+0.09}$ & $13.21_{-0.18}^{+0.16}$ & $6.21_{-0.21}^{+0.23}$ & $13.12_{-0.18}^{+0.16}$ & $0.99_{-0.00}^{+0.00}$ & $13.21_{-0.18}^{+0.16}$ & $0.00_{-0.00}^{+0.00}$ & $0.24_{-0.02}^{+0.02}$ & $0.13_{-0.00}^{+0.00}$ & $2.22_{-0.17}^{+0.18}$ & $0.00_{-0.00}^{+0.00}$\\
Hb1 & $10.21_{-0.45}^{+0.49}$ & $1.60_{-0.09}^{+0.09}$ & $11.45_{-0.03}^{+0.02}$ & $5.38_{-0.26}^{+0.28}$ & $11.38_{-0.02}^{+0.02}$ & $0.99_{-0.00}^{+0.00}$ & $12.26_{-0.59}^{+0.60}$ & $0.81_{-0.61}^{+0.63}$ & $0.15_{-0.01}^{+0.01}$ & $0.08_{-0.00}^{+0.00}$ & $1.38_{-0.09}^{+0.09}$ & $0.00_{-0.00}^{+0.00}$\\
\hline
\multicolumn{13}{c}{(15) PSR J1738+0333: $M_{\rm G} = 1.47_{-0.06}^{+0.07} M_\odot$, $\nu = 170.9$~Hz}\\
\hline
N1 & $10.32_{-0.38}^{+0.49}$ & $1.64_{-0.07}^{+0.09}$ & $11.57_{-0.04}^{+0.03}$ & $5.33_{-0.26}^{+0.24}$ & $11.50_{-0.03}^{+0.03}$ & $0.99_{-0.00}^{+0.00}$ & $12.52_{-0.51}^{+0.60}$ & $0.95_{-0.54}^{+0.64}$ & $0.15_{-0.01}^{+0.01}$ & $0.08_{-0.00}^{+0.00}$ & $1.44_{-0.08}^{+0.09}$ & $0.00_{-0.00}^{+0.00}$\\
N2 & $4.16_{-0.10}^{+0.12}$ & $1.60_{-0.07}^{+0.08}$ & $14.91_{-0.06}^{+0.06}$ & $6.87_{-0.28}^{+0.27}$ & $14.71_{-0.07}^{+0.08}$ & $0.99_{-0.00}^{+0.00}$ & $14.91_{-0.06}^{+0.06}$ & $0.00_{-0.00}^{+0.00}$ & $0.25_{-0.02}^{+0.02}$ & $0.13_{-0.00}^{+0.00}$ & $2.32_{-0.14}^{+0.16}$ & $0.00_{-0.00}^{+0.00}$\\
N3 & $5.73_{-0.23}^{+0.28}$ & $1.60_{-0.07}^{+0.08}$ & $14.49_{-0.04}^{+0.03}$ & $6.67_{-0.32}^{+0.30}$ & $14.32_{-0.03}^{+0.03}$ & $0.99_{-0.00}^{+0.00}$ & $14.49_{-0.04}^{+0.03}$ & $0.00_{-0.00}^{+0.00}$ & $0.22_{-0.01}^{+0.01}$ & $0.12_{-0.00}^{+0.00}$ & $2.08_{-0.11}^{+0.12}$ & $0.00_{-0.00}^{+0.00}$\\
Hp1 & $6.67_{-0.30}^{+0.40}$ & $1.62_{-0.07}^{+0.09}$ & $13.30_{-0.01}^{+0.00}$ & $6.12_{-0.28}^{+0.26}$ & $13.17_{-0.01}^{+0.00}$ & $0.99_{-0.00}^{+0.00}$ & $13.30_{-0.01}^{+0.00}$ & $0.00_{-0.00}^{+0.00}$ & $0.20_{-0.01}^{+0.01}$ & $0.10_{-0.00}^{+0.00}$ & $1.85_{-0.11}^{+0.12}$ & $0.00_{-0.00}^{+0.00}$\\
Hp2 & $5.77_{-0.27}^{+0.42}$ & $1.60_{-0.07}^{+0.08}$ & $14.49_{-0.05}^{+0.03}$ & $6.67_{-0.32}^{+0.30}$ & $14.32_{-0.04}^{+0.03}$ & $0.99_{-0.00}^{+0.00}$ & $14.49_{-0.05}^{+0.03}$ & $0.00_{-0.00}^{+0.00}$ & $0.22_{-0.01}^{+0.01}$ & $0.12_{-0.00}^{+0.00}$ & $2.08_{-0.11}^{+0.12}$ & $0.00_{-0.00}^{+0.00}$\\
S1 & $6.29_{-0.21}^{+0.28}$ & $1.82_{-0.08}^{+0.10}$ & $11.63_{-0.11}^{+0.12}$ & $5.35_{-0.19}^{+0.18}$ & $11.58_{-0.11}^{+0.12}$ & $1.00_{-0.00}^{+0.00}$ & $12.30_{-0.52}^{+0.60}$ & $0.67_{-0.41}^{+0.49}$ & $0.19_{-0.01}^{+0.01}$ & $0.10_{-0.00}^{+0.00}$ & $1.80_{-0.12}^{+0.13}$ & $0.00_{-0.00}^{+0.00}$\\
S2 & $3.92_{-0.09}^{+0.11}$ & $1.71_{-0.08}^{+0.09}$ & $13.27_{-0.14}^{+0.16}$ & $6.11_{-0.21}^{+0.19}$ & $13.19_{-0.14}^{+0.16}$ & $0.99_{-0.00}^{+0.00}$ & $13.27_{-0.14}^{+0.16}$ & $0.00_{-0.00}^{+0.00}$ & $0.25_{-0.02}^{+0.02}$ & $0.13_{-0.00}^{+0.00}$ & $2.29_{-0.15}^{+0.18}$ & $0.00_{-0.00}^{+0.00}$\\
Hb1 & $10.41_{-0.40}^{+0.51}$ & $1.64_{-0.08}^{+0.09}$ & $11.44_{-0.03}^{+0.02}$ & $5.27_{-0.26}^{+0.24}$ & $11.37_{-0.03}^{+0.02}$ & $0.99_{-0.00}^{+0.00}$ & $12.52_{-0.52}^{+0.61}$ & $1.08_{-0.54}^{+0.64}$ & $0.15_{-0.01}^{+0.01}$ & $0.08_{-0.00}^{+0.00}$ & $1.42_{-0.08}^{+0.09}$ & $0.00_{-0.00}^{+0.00}$\\
\hline
\multicolumn{13}{c}{(16) PSR J1918-0642: $M_{\rm G} = 1.18_{-0.09}^{+0.10} M_\odot$, $\nu = 131.6$~Hz}\\
\hline
N1 & $8.60_{-0.48}^{+0.56}$ & $1.28_{-0.11}^{+0.12}$ & $11.68_{-0.04}^{+0.03}$ & $6.70_{-0.54}^{+0.58}$ & $11.62_{-0.03}^{+0.03}$ & $0.99_{-0.00}^{+0.00}$ & $11.68_{-0.04}^{+0.03}$ & $0.00_{-0.00}^{+0.00}$ & $0.09_{-0.01}^{+0.01}$ & $0.07_{-0.00}^{+0.00}$ & $1.06_{-0.11}^{+0.12}$ & $0.00_{-0.00}^{+0.00}$\\
N2 & $3.72_{-0.13}^{+0.15}$ & $1.26_{-0.10}^{+0.12}$ & $14.54_{-0.10}^{+0.13}$ & $8.35_{-0.58}^{+0.63}$ & $14.42_{-0.14}^{+0.13}$ & $0.99_{-0.00}^{+0.00}$ & $14.54_{-0.10}^{+0.13}$ & $0.00_{-0.00}^{+0.00}$ & $0.14_{-0.02}^{+0.02}$ & $0.11_{-0.00}^{+0.00}$ & $1.65_{-0.19}^{+0.22}$ & $0.00_{-0.00}^{+0.00}$\\
N3 & $4.73_{-0.28}^{+0.33}$ & $1.26_{-0.10}^{+0.12}$ & $14.57_{-0.04}^{+0.03}$ & $8.35_{-0.67}^{+0.72}$ & $14.42_{-0.01}^{+0.03}$ & $0.99_{-0.00}^{+0.00}$ & $14.57_{-0.04}^{+0.03}$ & $0.00_{-0.00}^{+0.00}$ & $0.13_{-0.01}^{+0.01}$ & $0.11_{-0.01}^{+0.01}$ & $1.56_{-0.16}^{+0.17}$ & $0.00_{-0.00}^{+0.00}$\\
Hp1 & $5.53_{-0.25}^{+0.32}$ & $1.27_{-0.10}^{+0.12}$ & $13.20_{-0.04}^{+0.04}$ & $7.58_{-0.57}^{+0.60}$ & $13.11_{-0.04}^{+0.04}$ & $0.99_{-0.00}^{+0.00}$ & $13.20_{-0.04}^{+0.04}$ & $0.00_{-0.00}^{+0.00}$ & $0.11_{-0.01}^{+0.01}$ & $0.09_{-0.00}^{+0.00}$ & $1.34_{-0.15}^{+0.17}$ & $0.00_{-0.00}^{+0.00}$\\
Hp2 & $4.73_{-0.28}^{+0.33}$ & $1.26_{-0.10}^{+0.12}$ & $14.57_{-0.04}^{+0.03}$ & $8.36_{-0.68}^{+0.71}$ & $14.42_{-0.01}^{+0.03}$ & $0.99_{-0.00}^{+0.00}$ & $14.57_{-0.04}^{+0.03}$ & $0.00_{-0.00}^{+0.00}$ & $0.13_{-0.01}^{+0.01}$ & $0.11_{-0.01}^{+0.01}$ & $1.56_{-0.16}^{+0.17}$ & $0.00_{-0.00}^{+0.00}$\\
S1 & $5.44_{-0.21}^{+0.26}$ & $1.43_{-0.12}^{+0.13}$ & $11.01_{-0.24}^{+0.24}$ & $6.31_{-0.37}^{+0.38}$ & $10.98_{-0.23}^{+0.24}$ & $1.00_{-0.00}^{+0.00}$ & $11.01_{-0.24}^{+0.24}$ & $0.00_{-0.00}^{+0.00}$ & $0.10_{-0.01}^{+0.01}$ & $0.09_{-0.00}^{+0.00}$ & $1.27_{-0.15}^{+0.18}$ & $0.00_{-0.00}^{+0.00}$\\
S2 & $3.54_{-0.10}^{+0.12}$ & $1.35_{-0.11}^{+0.12}$ & $12.50_{-0.28}^{+0.28}$ & $7.17_{-0.40}^{+0.42}$ & $12.45_{-0.28}^{+0.28}$ & $1.00_{-0.00}^{+0.00}$ & $12.50_{-0.28}^{+0.28}$ & $0.00_{-0.00}^{+0.00}$ & $0.13_{-0.02}^{+0.02}$ & $0.11_{-0.00}^{+0.00}$ & $1.60_{-0.20}^{+0.22}$ & $0.00_{-0.00}^{+0.00}$\\
Hb1 & $8.66_{-0.47}^{+0.56}$ & $1.28_{-0.11}^{+0.12}$ & $11.49_{-0.01}^{+0.01}$ & $6.60_{-0.52}^{+0.55}$ & $11.44_{-0.01}^{+0.00}$ & $0.99_{-0.00}^{+0.00}$ & $11.49_{-0.01}^{+0.01}$ & $0.00_{-0.00}^{+0.00}$ & $0.09_{-0.01}^{+0.01}$ & $0.07_{-0.00}^{+0.00}$ & $1.04_{-0.11}^{+0.13}$ & $0.00_{-0.00}^{+0.00}$\\
\hline
\end{tabular}}
\end{table*}

\begin{table*}
\centering
\caption{Theoretically computed radius values of pulsars (Table~\ref{table_pulsar})
for non-spinning configurations, and the corresponding percentage bias in the 
allowed EoS models, if the EoS models are constrained using an accurate ($\pm 5$\%) equatorial radius
measurement (see \S~\ref{non-spinning}).}
\scalebox{0.80}{%
\hspace{-0.5cm}
\begin{tabular}{ccccccccccccccccccc}
\hline
\multicolumn{1}{c}{Pulsar\footnotemark[1]} & \multicolumn{1}{c}{$M_{\rm G}$\footnotemark[2]} & \multicolumn{1}{c}{$\nu$\footnotemark[3]} & \multicolumn{16}{c}{EoS models\footnotemark[4]}\\
no. & & & \multicolumn{2}{c}{N1} & \multicolumn{2}{c}{N2} & \multicolumn{2}{c}{N3} & \multicolumn{2}{c}{Hp1} & \multicolumn{2}{c}{Hp2} & \multicolumn{2}{c}{S1} & \multicolumn{2}{c}{S2} & \multicolumn{2}{c}{Hb1}\\
    & & & $R$\footnotemark[5] & Bias\footnotemark[6] & $R$ & Bias & $R$ & Bias & $R$ & Bias & $R$ & Bias & $R$ & Bias & $R$ & Bias & $R$ & Bias\\
    & ($M_\odot$) & (Hz) & (km) & (\%) & (km) & (\%) & (km) & (\%) & (km) & (\%) & (km) & (\%) & (km) & (\%) & (km) & (\%) & (km) & (\%)\\ 
\hline 
1 & 1.667 & 465.1 & 11.41 & 20.0 & 14.99 & 47.1 & 14.27 & 45.1 & 13.18 & 33.8 & 14.22 & 47.1 & 11.90 & 19.9 & 13.68 & 31.6 & 11.29 & 19.4\\
2 & 1.41 & 416.7 & 11.56 & 20.4 & 14.76 & 43.0 & 14.42 & 44.5 & 13.23 & 31.0 & 14.42 & 44.5 & 11.50 & 15.8 & 13.12 & 25.6 & 11.42 & 19.3\\
3 & 1.438 & 366.0 & 11.55 & 15.3 & 14.79 & 31.7 & 14.41 & 32.4 & 13.23 & 23.3 & 14.41 & 32.4 & 11.55 & 10.9 & 13.18 & 18.4 & 11.41 & 14.4\\
4 & 1.540 & 339.0 & 11.49 & 12.0 & 14.89 & 24.8 & 14.36 & 24.7 & 13.23 & 18.8 & 14.35 & 25.1 & 11.73 & 9.3 & 13.42 & 14.9 & 11.37 & 11.4\\
5 & 1.928 & 317.5 & 11.11 & 7.8 & 15.13 & 19.5 & 14.01 & 17.0 & 12.82 & 14.4 & 13.32 & 25.3 & 12.07 & 8.8 & 14.10 & 11.9 & 10.99 & 7.8\\
6 & 1.832 & 315.5 & 11.24 & 8.1 & 15.08 & 20.1 & 14.12 & 17.8 & 13.02 & 14.0 & 13.84 & 20.7 & 12.05 & 8.0 & 13.96 & 11.7 & 11.14 & 7.9\\
7 & 1.33 & 305.8 & 11.60 & 11.6 & 14.67 & 23.4 & 14.45 & 24.7 & 13.21 & 17.2 & 14.45 & 24.7 & 11.34 & 6.8 & 12.91 & 11.7 & 11.44 & 11.0\\
8 & 1.64 & 287.4 & 11.43 & 7.6 & 14.97 & 16.5 & 14.29 & 16.5 & 13.20 & 12.6 & 14.25 & 17.3 & 11.87 & 5.9 & 13.62 & 9.6 & 11.31 & 7.6\\
9 & 1.393 & 279.3 & 11.57 & 9.3 & 14.74 & 18.2 & 14.42 & 18.8 & 13.23 & 13.9 & 14.42 & 18.8 & 11.47 & 5.4 & 13.07 & 9.5 & 11.43 & 8.7\\
10 & 1.366 & 238.7 & 11.58 & 6.9 & 14.71 & 13.2 & 14.43 & 13.7 & 13.22 & 10.3 & 14.43 & 13.7 & 11.42 & 1.9 & 13.00 & 5.7 & 11.43 & 6.6\\
11 & 1.33 & 218.8 & 11.60 & 6.1 & 14.67 & 11.3 & 14.45 & 11.9 & 13.21 & 8.8 & 14.45 & 11.8 & 11.34 & 1.6 & 12.91 & 1.9 & 11.44 & 5.7\\
12 & 1.83 & 190.1 & 11.24 & 3.4 & 15.08 & 8.6 & 14.13 & 6.8 & 13.02 & 5.3 & 13.85 & 8.0 & 12.05 & 2.1 & 13.96 & 2.2 & 11.14 & 3.3\\
13 & 1.30 & 186.6 & 11.61 & 4.6 & 14.63 & 8.3 & 14.46 & 8.8 & 13.20 & 6.6 & 14.46 & 8.8 & 11.28 & 1.1 & 12.83 & 1.3 & 11.45 & 4.4\\
14 & 1.44 & 173.6 & 11.55 & 3.8 & 14.79 & 6.7 & 14.41 & 7.0 & 13.23 & 5.4 & 14.40 & 7.0 & 11.56 & 1.1 & 13.18 & 1.7 & 11.41 & 3.4\\
15 & 1.47 & 170.9 & 11.53 & 3.5 & 14.82 & 6.1 & 14.39 & 6.7 & 13.24 & 5.1 & 14.39 & 6.6 & 11.61 & 1.1 & 13.26 & 1.3 & 11.40 & 3.3\\
16 & 1.18 & 131.6 & 11.65 & 2.7 & 14.49 & 4.1 & 14.49 & 5.5 & 13.15 & 3.6 & 14.49 & 5.6 & 11.00 & 0.5 & 12.49 & 0.6 & 11.46 & 2.5\\
\hline
\end{tabular}}
\begin{flushleft}
$^1$Pulsar numbers from Table~\ref{table_pulsar}.\\
$^2$Gravitational mass of pulsars.\\
$^3$Spin frequency of pulsars.\\
$^4$Equation of state models (Table~\ref{table_EoS} and Fig.~\ref{fig1}).\\
$^5$Radii of pulsars for non-spinning configurations.\\
$^6$This quantity gives the percentage bias in the allowed
EoS models, if the EoS models are constrained using a stellar equatorial
radius measured with $\pm 5$\% accuracy and using theoretical non-spinning configurations (see \S~\ref{non-spinning}).
\end{flushleft}
\label{table_diff}
\end{table*}

\clearpage

\label{lastpage}
\end{document}